\documentclass[a4paper,fleqn]{cas-sc}
\usepackage{float,graphicx,amsmath,multirow}
\usepackage{color}
\usepackage{siunitx}
\usepackage[version=3]{mhchem}
\usepackage{rotating}
\usepackage[authoryear]{natbib}

\def\tsc#1{\csdef{#1}{\textsc{\lowercase{#1}}\xspace}}
\tsc{WGM}
\tsc{QE}
\begin{document}
\let\WriteBookmarks\relax
\def\floatpagepagefraction{1}
\def\textpagefraction{.001}

\shorttitle{Ice Chemistry Modeling of Comet Hale--Bopp}    

\shortauthors{Willis, Christianson, and Garrod}  

\title [mode = title]{Ice Chemistry Modeling of Active Phase Comets: Hale--Bopp}

\author[1]{Eric R. Willis}

\affiliation[1]{organization={Department of Chemistry, University of Virginia},
            addressline={409 McCormick Rd}, 
            city={Charlottesville},
            postcode={22904}, 
            state={Virginia},
            country={USA}}

\author[1]{Drew A. Christianson}

\cormark[1]

\author[1,2]{Robin T. Garrod}

\affiliation[2]{organization={Department of Astronomy, University of Virginia},
            addressline={530 McCormick Rd}, 
            city={Charlottesville},
            postcode={22904}, 
            state={Virginia},
            country={USA}}

\cortext[1]{Corresponding author}

\begin{abstract}
We present a chemical kinetics model of the solid-phase chemical evolution of a comet, beginning with a long period of cold-storage in the Oort Cloud, followed by five orbits that bring the comet close to the Sun. Current orbital parameters for Hale--Bopp are used for all orbits. The chemical model is based on an earlier treatment (Garrod 2019) that considered only the cold-storage phase, and which was based on the interstellar ice chemical kinetics model \emph{MAGICKAL}. The comet is treated as 25 chemically distinct layers, whose thicknesses increase with depth; each layer has a dust component, and the chemical network includes a further 200 atomic/molecular species. The new model includes several key updates to the previous treatment: (i)~Time- and depth-dependent temperature profiles based on heat transfer according to heliocentric distance; (ii)~a rigorous treatment of back-diffusion (random walk) for species capable of diffusing through the thick bulk-ice layers; (iii)~adoption of recent improvements in the kinetic treatment of nondiffusive chemical reaction rates in the ice and on the ice surface. Starting from an initially simple ice composition, interstellar UV photons drive a rapid chemistry in the upper micron of material, but diminished by absorption of the UV by the dust component. Galactic cosmic rays (GCRs) drive a much slower chemistry in the deeper ices over the long cold-storage period, to depths on the order of 10~m. The first solar approach drives off the upper layers of ice material via thermal desorption and/or dissociation (the model does not include a treatment for the more complex cometary outbursts), bringing closer to the surface the deeper material that previously underwent long-term processing by GCRs. Subsequent orbits are more uniform in their chemical behavior after the upper layers are lost. Loss of molecular material leads to concentration of the dust in the upper layers, with a large dust fraction extending to depths on the order of 10~cm after all orbits are complete. Substantial quantities of complex organic molecules (COMs) are formed in the upper 10~m during the cold storage phase, with some of this material released during solar approach; however, their abundances with respect to water appear too low to account for the observed gas-phase values for comet Hale--Bopp, indicating that the majority of complex molecular material observed, at least in comet Hale--Bopp, is an inheritance of primordial material.
\end{abstract}

% Keywords

\maketitle

% Main text
\section{Introduction}
\label{intro}

Cometary nuclei are believed to contain some of the most primitive and well-preserved material from the formation of the Solar System \citep{Weissman2020}. Generally, comets are stored in one of two regions of the Solar System -- the Kuiper Belt \citep{Kuiper1951} and the Oort cloud \citep{1950BAN....11...91O}; from here, some gravitational perturbation may place a comet into an orbit that more closely approaches the Sun. The Kuiper Belt extends from the orbit of Neptune ($\sim$30 AU) out to $\sim$50 AU, and has a mean temperature of $\sim$40~K \citep{Hsiah2006}. The Oort cloud is significantly larger and colder, with suggested inner and outer radii of 3000 and 50,000~AU, respectively, and a mean temperature of $\sim$10~K \citep{Hsiah2006}. Other estimates have placed the maximum radius of the Oort cloud at a value as large as 200,000~AU \citep{Duncan1987}. 

Cometary ice is dominated by \ce{H2O}, yet complex organic molecules (COMs) -- typically defined as organic species with 6 or more atoms -- have also been detected in comets via remote observations, as well as from returned samples \citep{Mumma2011}. The long-period comet C1/1995 (Hale--Bopp), which has an Oort cloud origin, was found to harbor an array of complex organics. \citet{Bockelee-Morvan2000} detected formamide (\ce{NH2CHO}) and methyl formate (\ce{HCOOCH3}) in the coma of Hale--Bopp using ground-based telescopes. In the same comet, \citet{Crovisier2004a} detected ethylene glycol (\ce{(CH2OH)2}), with a production rate from the cometary nucleus of around 0.25\% that of \ce{H2O}, indicating a substantial quantity of this molecule in the ice. Observations of Hale--Bopp beginning in August 1995 detected outgassing of nine other molecular species, including methanol (\ce{CH3OH}), formaldehyde (\ce{H2CO}), and acetonitrile (\ce{CH3CN}) \citep{Biver1997}. Although the abundance of methanol -- usually considered the simplest COM -- that was detected toward Hale--Bopp ($\sim$2.4\% with respect to water) is not unusually high compared with other Oort Cloud comets, the brightness of Hale--Bopp allowed for the observational detection of an unusually broad range of larger, less abundant COMs. This great number of molecular data-points thus makes Hale--Bopp a good candidate for the modeling of possible COM chemistry in the comet nucleus.

Samples returned from the \textit{Stardust} mission to comet 81P/Wild 2 \citep{Brownlee2006} have also revealed significant chemical complexity. \citet{Glavin2008} utilized a combination of liquid chromatography and UV fluorescence to make detections of methylamine (\ce{CH3NH2}) and ethylamine (\ce{CH3CH2NH2}) in the returned samples. Following on from that study, \citet{Elsila2009} studied the \ce{^{13}C} content of amino acids in the solid samples returned from Wild 2, reporting the first detection of a cometary amino acid, glycine (\ce{NH2CH2COOH}).

More recently, \citet{Altwegg2016} used the ROSINA mass spectrometer \citep{Balsiger2007} to detect glycine, methylamine, and ethylamine in the coma of Comet 67P/Churyumov-Gerasimenko. These detections, along with the simultaneous identifications of phosphorus-bearing molecules and other organics, have confirmed the discoveries made by the \textit{Stardust} team, and highlighted the role that comets could have played in seeding the early Earth with organic molecular building blocks. 

Many sulfur-bearing molecules have also been detected in the coma of Comet 67P/Churyumov-Gerasimenko, such as \ce{H2S}, \ce{SO}, \ce{S2}, \ce{S3}, \ce{CH3SH}, and \ce{C2H5SH}. Some sulfur-bearing species are used as bio-signatures in the search for life, and the presence of species like \ce{S2} in Comet 67P indicates the ice mantles formed during or prior to the solar nebula and were, at least in part, maintained to this day, as opposed to being formed or reformed later in the solar systems lifetime \citep{Calmonte2016}. Additionally, many ammonium salts have been detected in cometary dust outbursts \citep{Altwegg2020} including more recently detections of \ce{NH4+SH-} by \citet{Altwegg2022} within the cometary dust specifically, while remaining more independently as \ce{NH3} and \ce{H2S} within the cometary ice, both of which are some of the most abundant species within the ice. 

Understanding the origin and evolution of COMs in cometary ices is of fundamental importance. Comets are thought to preserve much of their molecular content from the proto-solar nebula stage, and perhaps even earlier; \citet{Drozdovskaya2019} compared the composition of comet 67P with the abundances of gas-phase molecules detected toward the low-mass star forming region IRAS 16293-2422, finding correlations between their N-, S- and HCO-group-bearing species. However, the degree of processing that cometary ices may have undergone during later stages of evolution, including as fully-formed comets, is difficult to discern. The ongoing development of detailed chemical kinetic models of cometary chemistry provides an opportunity to simulate the production, survival and/or destruction of molecules -- including COMs -- at these later stages.

Models of gas-phase chemistry in cometary comae have existed for some time \citep[e.g.][]{Irvine1998}. Those models only included gas-phase processes in active comets, and were adapted from models constructed to simulate interstellar chemistry. One of the key questions motivating their use has been in distinguishing between those species formed in the cometary ice (``parent'' species) and those formed via energetic processing in the coma (``daughter'' species). However, in order to properly determine which species are produced in the coma, and which are inherited directly from the cometary nucleus, detailed models of the solid-phase chemistry in comets are needed.

\citet{Garrod2019} (hereafter G19) published the first such computational model of solid-phase chemistry in cometary ice, which used an adapted version of the interstellar astrochemical code \emph{MAGICKAL} \citep[\textit{Model for Astrophysical Gas and Ice Chemical Kinetics and Layering;}][]{Garrod2013b}. In the model of G19, the comet nucleus was constructed by extending the grain-surface and ice-mantle chemistry already included in \emph{MAGICKAL}, while switching off the gas-phase chemistry. Although photochemistry was already present, new reactions were added to the network to take account of molecular dissociation induced by cosmic ray impacts.  The nucleus material consisted of both ice material composed of various molecules and atoms, and a variable dust component that was chemically inert but capable of diminishing the impinging interstellar UV field. The physical structure of the comet was modeled by dividing the nucleus into 25 layers, each three times thicker than the last, beginning with a 1 monolayer-thick outer layer. The total depth into the nucleus was $\sim$136~m. Chemistry was studied for the ``cold storage'' phase of an Oort cloud comet's lifecycle, before the onset of any activity caused by solar approach. Significant COM formation was observed to a depth of $\sim$15~m, due to processing of the ice by galactic cosmic rays, with the upper $\sim$1$\mu$m affected also by the penetration of UV photons.

Experimental work by \citet{Gronoff2020} and \citet{Maggiolo2020} also examined the effects of GCRs on cometary ices. While the effect of GCRs on isotopic ratios are negligible, the chemical processing is significant down to depths of tens of meters, although a more active collisional history between comets may lead to mixing of ices, allowing for processed, non-pristine, material to exist at greater depths. However, if a comet has no collisions, or even just a few soon after formation, then the ice should remain pristine below the tens of meters mark. 

Furthermore, the material outgassed by a comet, either within a localized outburst or by sublimation, is highly dependent on its dynamical history. A comet on its first close solar approach will eject material that has been processed by GCRs, while much older comets that have had their surfaces significantly eroded away may instead eject more pristine ice. Both in intermediary stages, as well as in cases with more recent cometary collisions, it is likely to see some mixture of processed and pristine ice within outbursts which may be localized to certain parts of the comet. 

Although comets spend the majority of their time in cold storage, either in the Kuiper Belt or the Oort Cloud, the solar approach of cometary bodies is very interesting from a chemical perspective. As noted above, cometary bodies eject material from their surfaces when approaching the Sun, and COMs have been detected in the resulting coma material. However, there have been no models of solid-phase cometary chemistry that consider solar approach.

In this paper, we present an updated version of the cometary ice chemistry model of G19, in which we consider both the long, ``cold-storage'' phase in the Oort Cloud and the subsequent solar approach, following the orbital dynamics of comet C1/1995 (Hale--Bopp) as established in the literature. Based on distance-dependent solar irradiation, we solve the heat diffusion equation in one dimension, allowing time- and depth-dependent temperature profiles to be obtained, which are then incorporated into the chemical model. The model now incorporates a full range of non-thermal chemical reaction mechanisms \citep{JG20,G22}, which allows bulk-ice reactions to be treated accurately, including reactions initiated by photodissociation or CR-induced dissociation of molecules. The model also includes a new treatment to take account of the back-diffusion effect, when calculating rates of bulk diffusion for specific chemical species between ice layers of different depths and thicknesses. 

\S \ref{methods} describes the new mechanisms incorporated into the chemical model, as well as the heat transfer and back-diffusion simulations. \S \ref{results} presents the results of the heat-transfer and back-diffusion simulations, as well as our chemical model results. \S \ref{discussion} compares the results from these models with those of G19, as well as observations of Hale--Bopp. Conclusions follow in \S \ref{conclusions}.

\section{Methods}
\label{methods}

Here we build upon the model of ice chemistry in cometary nuclei introduced by G19. The initial model was adapted from the hot-core chemical kinetics model, \emph{MAGICKAL} \citep{Garrod2013b}. The reader is referred to G19 for a full description of the comet chemical model. Below we outline basic elements of the model, along with the major updates that have been made to this new version.

The new comet model moves beyond the pure cold-storage treatment studied by G19, and presents several improvements. Firstly, we have incorporated a range of new non-diffusive ice chemistry mechanisms, following \citet{JG20} and \citet{G22} (see \S \ref{nonthermal_meth}). A more accurate treatment of back-diffusion between ice layers within the comet, building on the work of \citet{Willis2017}, is also included in the model (\S \ref{back_meth}). In the models presented here, the cold-storage phase is followed by the solar approach of the comet (\S \ref{heat_meth}).

\subsection{Basic model functions}
\label{basic_meth}

The cometary solid-phase chemical kinetics model of G19 divided the comet into 25 distinct layers of ice/dust, each of thickness three times greater than the last, beginning with an upper layer one monolayer thick. These 25 layers collectively produced a thickness corresponding to $\sim$136~m; beneath this was an inert reservoir layer of indefinite thickness that would replenish any shortfall in the above layers.

Within each of the 25 regular ice/dust layers, the behavior of a set of 
chemical species was traced over time. The chemistry occurring in each layer included diffusive reactions between various species, photodissociation caused by Galactic UV photons, dissociation caused by Galactic cosmic-ray impingement, spontaneous reactions (i.e. ``non-diffusive'' reactions, see below) between newly-produced dissociation products and pre-existing local reaction partners, and -- in the surface layer only -- the thermal and non-thermal desorption of species according to their binding energies and the temperature of the comet. The temperature in the G19 models was assumed to be uniform through all layers and at all times in each model run. No adsorption of atoms or molecules from the gas-phase onto the surface was allowed, based on the assumption of low local gas densities, and this assumption is retained in the present model.

\begin{figure}[h!]
    \centering
    \includegraphics[width=0.995\textwidth]{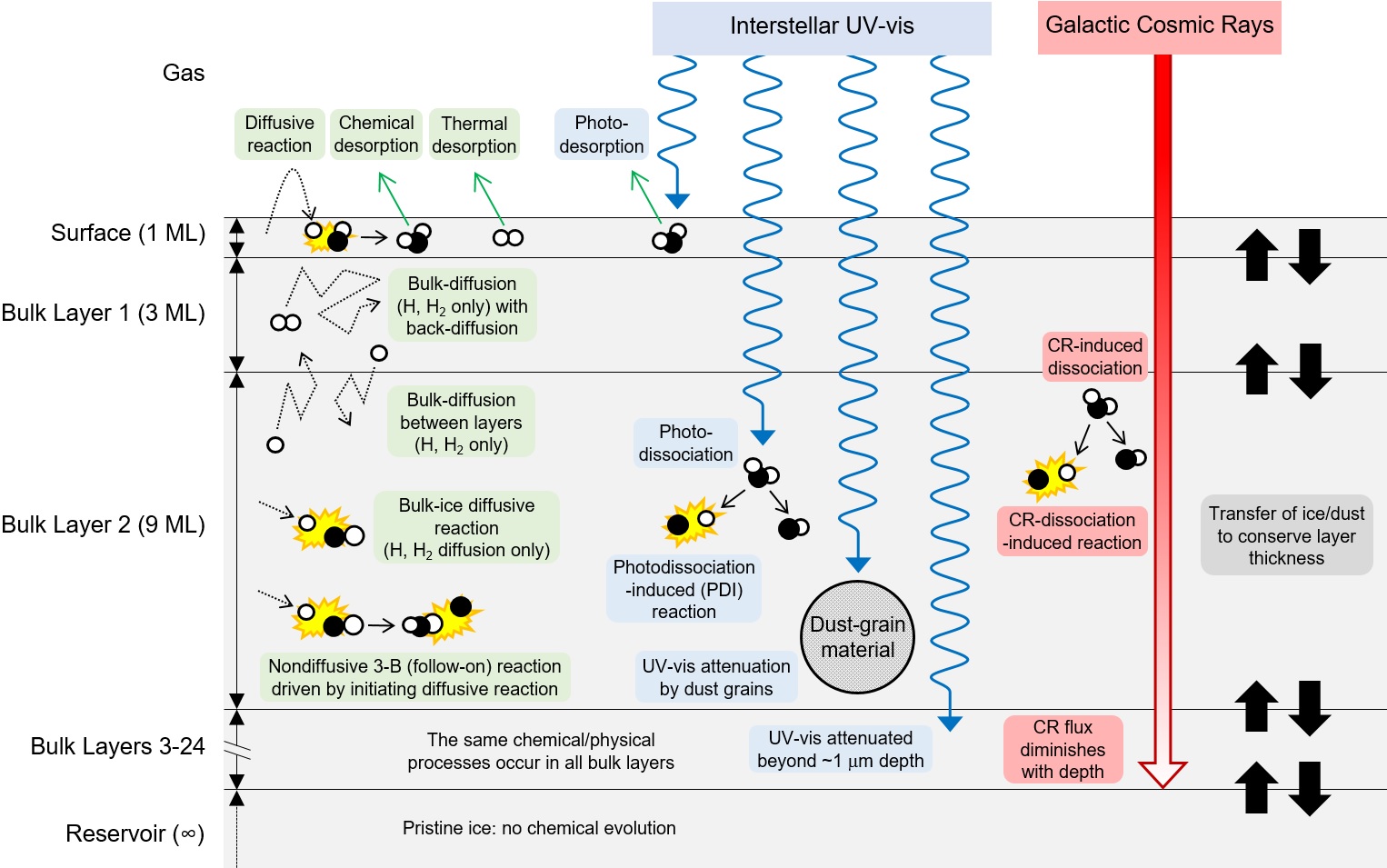}
    \caption{A simplified pictorial schematic of some of the physical processes modeled by \emph{MAGICKAL}. Each bulk (i.e. sub-surface) layer is three times thicker than the last. Dotted arrows indicate surface or bulk-ice diffusion. Solid black arrows indicate the products of reaction or dissociation. Each ice/dust layer is assigned its own local temperature, while molecular dissociation rates (induced by UV--Vis photons and Galactic cosmic rays) are also locally determined. The same chemical reactions and physical processes are allowed in the surface layer and all bulk layers, except for desorption processes, which only occur in the surface layer. Aside from H and H$_2$, molecular shapes shown are intended only to be representative of generic structures. Some processes have been omitted for clarity.}
    \label{mag_scheme}
\end{figure}

Transfer of chemical species between the 25 layers was allowed in the G19 model, caused by diffusion of mobile species across layer boundaries. Transfer was also allowed as the result of losses induced in the upper layer (due to desorption), or of either losses or gains in the total amount of ice/dust in any of the layers, as the result of the chemistry. This transfer was monitored throughout the model, to ensure that the amount of material in all layers was held constant over time. Any additional material required in the deepest (25$^{th}$) layer was drawn from the reservoir beneath, which was assumed to retain a pristine ice composition throughout (based on the chosen starting composition in the model). Excess material in the deepest layer was likewise transferred into the reservoir. Figure~\ref{mag_scheme} shows a simplified schematic of how the updated version of the \emph{MAGICKAL} comet model presented here functions.

The chemical composition of the comet was assumed to include not only ice but also a dust component. Although the dust was treated as chemically inert, it was allowed to undergo transfer between layers according to losses or gains in other species; in this way, desorption of volatile material from the surface of the comet would lead to the enrichment of the dust component of the upper layers, as material (both ice and dust) from the next layer down replaced the lost material from the surface. The effect of the attenuation of the UV photon field by the dust was also considered, meaning that the dust enrichment of the upper layers would also serve to reduce somewhat the rates of UV photodissociation in the layers beneath. 

Although the model has several mechanisms for the loss of atoms/molecules from the surface layer (e.g.~thermal sublimation), it does not include any means for the ejection of dust from the surface. Dust loss from a comet could occur either through local outburst events or as the result of sublimation of icy molecular material; however, the inclusion of rates for these mechanical dust-loss processes in the chemical model is non-trivial and is thus left for future investigation.

Here we make a point of clarifying our definitions of two important mass-loss processes: (i) outbursts, which are characterized by a sudden and brief increase in brightness sometimes attributed to the sudden local ejection of dust and trapped volatiles, and (ii) outgassing, which is a continuous process of sublimation of ices that is dependent on heliocentric distance via temperature. While our model does not include the localized and brief events associated with cometary outbursts, it does consider outgassing through the continuous process of sublimation (although, as noted above, without any concomitant dust loss).

The initial fractional composition of the comet was set to be the same as in G19. As water is the most abundant ice species, other fractional abundances are set with respect to water ice. \ce{CO} and \ce{CO2} are each set to 20\%, Methanol is set to 5\%, and methane, formaldehyde, and ammonia are each set to 1\%. These fractional abundances are based on the dominant components of interstellar ices. We also start with an initial dust content equal to 111\% with respect to water ice. These initial ice fractions apply to the surface layer as well as all bulk ice layers including the reservoir. As the chemical model progresses, we also allow a maximum dust content per layer of $\sim$74\% of the total composition. This value is based on the maximum space-filling factor for identically sized spheres \citep[e.g.][]{Hales2005}. 

The G19 chemical models were applied only to static physical conditions intended to represent the cold-storage phase of cometary evolution. The present models include both cold storage and a solar approach phase, described in detail below.

Here we also employ an expanded chemical network based on that of \citet{Willis2020}, with all species that contain more than 5 carbon atoms removed to make the integration of the chemical equations more easily manageable.

Other changes to the modeling treatment of G19 as applied to the present model are described below. A list of reactions with activation energies and barrier widths used in the model is available online.\footnote{https://garrodgroup.as.virginia.edu/resources}

\subsection{Non-diffusive chemical mechanisms}
\label{nonthermal_meth}

In most prior gas-grain astrochemical kinetics models, the production of COMs has relied on diffusive chemistry occurring between functional-group radicals on interstellar dust-grain surfaces or within the ices that build up on those surfaces \citep[e.g.,][]{Hasegawa1992,Garrod08,Garrod2013}. These mechanisms generally require the dust grains to reach temperatures at which larger radical species can become mobile, which may easily be reached in the star-forming cores where many COMs are found. However, recent gas-phase detections of COMs in cold environments \citep[e.g.,][]{Bacmann2012,Cernicharo2012,Bergner2017a} indicate that, if such molecules are indeed formed on the grains, then formation mechanisms must be operative that can occur at much lower temperatures, at which large radicals remain thermally immobile. The implications of these findings extend also to cometary ice chemistry, as comets in the Oort cloud exist at very low temperatures ($\sim$10~K) for very long time periods.

In the comet model presented by G19, chemistry was assumed to occur both through diffusive meetings of reactants, as well as through a novel, non-diffusive photodissociation-induced mechanism, whereby reactive radicals produced by photolysis in the ice could spontaneously react with nearby reaction partners. A similar non-diffusive mechanism was also included to account for reactions induced by direct cosmic-ray impingement, following the production of reactive radicals deep within the ice via radiolysis.

Subsequent interstellar gas-grain chemistry models by \cite{JG20} and \cite{G22} further refined the formulation of the non-diffusive photodissociation-induced reactions, and generalized it to allow similar formulations for other non-diffusive reaction processes. The new comet models presented here incorporate these and other recent advances in surface/solid-phase interstellar chemistry.

Firstly, the ``photo-dissociation induced'' (PDI) mechanism of non-diffusive reactions has been updated as described by \cite{JG20} and \cite{G22}; the latter authors took into consideration the efficiency at which photo-products would recombine with each other if no other reaction partner were immediately present. The effect of this change was to reduce somewhat the production of unreacted radicals in the bulk ice. All changes to the photon-induced mechanisms are mirrored precisely in the corresponding cosmic-ray induced processes.

An additional non-diffusive chemical mechanism is added to the model: the so-called ``three-body'' (3-B) reaction mechanism, described in detail by \cite{JG20} and \cite{G22}. In this mechanism, a preceding surface or bulk-ice reaction may lead to a follow-on reaction if some third reactive body happens to be in close proximity to the site of the first reaction. The follow-on reaction could involve an activation energy barrier, and the follow-on reaction therefore need not occur immediately following the initiating reaction, so long as the reactants remain undisturbed for long enough for the barrier to be eventually overcome. The initiating reaction could be a diffusive process, a PDI (or similar cosmic ray-induced) process, or another three-body reaction. Following \citet{G22}, we allow a total of three cycles of 3-B reactions to occur. Within the scheme of the 3-B process, our model also includes adjustments to permit a sufficiently exothermic initiating reaction to allow the activation energy barrier of the follow-on reaction to be instantaneously overcome, as described by \cite{JG20} and formulated by \cite{G22}. A form of the 3-B process was initially included in the gas-grain model of \citet{Garrod2011} that was used to explain the production of interstellar solid-phase CO$_2$.

The generic form for calculating the rates of non-diffusive processes, between two species A and B, is given by

\begin{equation}
    \label{generic}
    \begin{aligned}
    R_{AB} = f_{act}(AB)R_{comp}(A) \frac{N(B)}{N_M} \\ + f_{act}(AB)R_{comp}(B) \frac{N(A)}{N_M},
    \end{aligned}
\end{equation}
where $f_{act}$ is an efficiency related to the activation energy barrier (between 0 and 1), $R_{comp}$ is the ``completion rate'' for species A or B, $N(A)$ and $N(B)$ are the abundances of species A and B in a particular layer, and $N_{M}$ is the total abundance of all species in that layer \citep[see][]{JG20}.

The ``completion rate'' of some species A, involved in reaction with species B, depends on two quantities; the appearance rate of A (for example, the total rate at which species A is produced by photodissociation, in the case of a PDI process), along with an adjustment that takes account of any delay in the follow-on reaction related to the presence of an activation energy barrier. Again, \citet{JG20} provides a detailed description of the formulation of the non-diffusive reaction rates.

Finally, following \citet{G22}, diffusion within the bulk ice in the new comet models is prohibited for species other than \ce{H} and \ce{H2} (surface diffusion by all species is nevertheless allowed as before). This is in line with experimental and modeling evidence \citep[e.g.][]{Ghesquiere2018,Shingledecker2019b} that indicates that heavier radicals do not diffuse rapidly, and as such that bulk diffusion may not be the main mechanism by which bulk-ice chemistry is driven. It is assumed here that bulk diffusion by H/\ce{H2} involves movement between interstitial sites in the ice matrix \citep{Chang14}. In the present comet models, this restriction means that diffusive reactions in the bulk ice may only occur if they involve either \ce{H} or \ce{H2}. Furthermore, diffusion between bulk-ice layers is also restricted to those two species. The outcome of this change on the interstellar chemical model results is described by \cite{G22}.

It should be noted that all of the chemical reactions included in the network may occur through any of the diffusive or non-diffusive mechanisms used in the models, so long as one or other of the reactants is allowed to diffuse (according to the above stipulation), or that one or other reactant may be formed by some preceding mechanism (e.g. photodissociation, or another reaction), in the case of non-diffusive processes.
\\

\subsection{Back-diffusion between layers}
\label{back_meth}

Back-diffusion is the phenomenon by which particles diffusing in/upon a lattice undergo a random walk, such that they may re-visit lattice sites that they have already diffused away from. This effect has been shown to impact reaction rates on the surfaces of interstellar dust grains of various morphologies. In the study of back-diffusion on interstellar dust grains by \citet{Willis2017}, the authors used kinetic Monte Carlo models to demonstrate that the surface reaction rate for the simple chemical system of H + H $\rightarrow$ \ce{H2} is decreased when back-diffusion is considered, compared with the typical rate-equation chemical kinetics approach. The magnitude of this effect was observed to be inversely proportional to the surface coverage of the diffusing particle. Thus, when the surface was mostly covered with diffusers, the effect was negligible, whereas with just two diffusers a maximum back-diffusion factor of $\sim$5 was observed, diminishing the reaction rates by this factor.

While the latter study considered the effect of back-diffusion on diffusive chemical reaction rates on grain surfaces, similar effects in three dimensions may affect both the rates of diffusive reactions in the bulk-ice layers and the rates at which diffusers may pass between the layers; the comet model includes terms that allow diffusers in one layer to pass upward or downward to an adjacent layer. As discussed above, in the models presented here, diffusion within the bulk ice is limited to H and \ce{H2} and is assumed to involve movement between interstitial sites within the ice structure.

In the comet models of G19, the rate of bulk diffusion at which some species $i$ diffuses into the surface layer from the thicker mantle layer below, $R_{\mathrm{diff},m}(i)$, was defined according to
\begin{equation}
    \label{rdiff}
    R_{\mathrm{diff},m}(i) = k_{\mathrm{diff}}(i) N_{m}(i) \frac{N_{S}}{6N_{M}}.
\end{equation}
\begin{equation}
    \label{kdiff}
    k_{\mathrm{diff}}(i) = \nu_{0}(i) \text{exp}[-E_{\mathrm{diff}}(i)/T],
\end{equation}
where $N_{m}(i)$ corresponds to the population of species $i$ within a particular mantle layer, $N_{S}$ and $N_{M}$ are the total number of available sites that can be occupied in the surface and bulk, respectively, $k_{\mathrm{diff}}(i)$ is the diffusion rate coefficient, $\nu_{0}(i)$ is the characteristic frequency of species $i$, $E_{\mathrm{diff}}(i)$ is the barrier to bulk diffusion of species $i$, and $T$ is the temperature of the ice layer. 

G19, following  \citet{Garrod2013b}, considered bulk diffusion to be a ``swapping'' process, but the same formulation is applicable to the present assumption that diffusion of H and \ce{H2} occurs between interstitial sites. Similar expressions were used by G19 to describe the rates of diffusion across the interfaces of all 25 layers in the comet model.

Eq.~(\ref{rdiff}) describes a situation in which there is a high degree of occupation by diffusers of the available sites in the ice. Thus, there is always a substantial number of diffusers directly adjacent to the interface between layers, such that any one of them might cross with a single ``hop'' (i.e.~diffusion event); the overall rate of diffusion across the interface is therefore dominated by the diffusion rates of those interface-adjacent diffusers. Eq.~(\ref{rdiff}) is composed of: the total rate at which any diffuser of type $i$ in the ice layer makes a single hop in any direction, given by $k_{\mathrm{diff}}(i) N_{m}(i)$; the fraction of all diffusers that are adjacent to the interface, $N_{S} / N_{M}$; and a factor 6 that comes from the six diffusion directions available to a particle at the interface between the surface and the mantle ice layer, on the assumption of a cubic ice structure.

Unfortunately, since the abundances of the diffusing H and \ce{H2} particles in the comet models could fall far short of maximum occupation, back-diffusion could have a significant impact on the rates of transfer between layers, with diffusers having to undergo multiple hops within the bulk before reaching the interface, including many hops that might lead them away from the interface in question. Eq.~(\ref{rdiff}) in that case could be highly inaccurate.

\citet{Garrod2017} considered such a situation in the context of bulk-diffusion in interstellar dust-grain ice mantles, which involves ices with a maximum thickness around a few hundred monolayers. Those authors used a simple three-dimensional kinetic Monte Carlo model to determine the average number of diffusion events (denoted by $N_{\mathrm{move}}$) required for a lone diffuser to reach the top layer of a monolithic ice mantle of finite thickness from some arbitrary starting point. (Note that escape from the lowest monolayer is blocked by the presence of the dust grain itself). The following relationship was found to provide an excellent fit to the data:
\begin{equation}
    \label{singlediff}
    N_{\mathrm{move}} = 2(N_{\mathrm{th}} + 1/2)^2,
\end{equation}
where $N_{\mathrm{th}}$ is the thickness of the ice in monolayers (equal to $N_{M} / N_{S}$). This expression indicates that, with increasing thickness, the number of diffusion events needed to reach the interface between the surface and the mantle increases rapidly. Eq.~(\ref{singlediff}) describes an extreme case, in which there is only one diffuser present. In the other extreme case, provided by Eq.~(\ref{rdiff}), in which diffusers are very abundant, $N_{\mathrm{move}} = 6 N_{\mathrm{th}} = 6 N_{M} / N_{S}$.

As noted by \citet{Willis2017} in the context of surface diffusion, the strength of the back-diffusion effect is dependent on the abundance of the diffusers. Furthermore, the degree of back-diffusion indicated by Eq. (\ref{singlediff}) should become extremely large when dealing with ice layers of macroscopic size, leading to a large divergence between the two extreme cases. In order to provide an accurate picture of diffusion between the layers in the comet model, a treatment is thus required that can account for back-diffusion anywhere between the two extreme cases corresponding to ``many'' or ``few'' diffusers.

As part of this paper, we present the results of a three-dimensional Monte Carlo (MC) model that explicitly simulates the back-diffusion effect in a bulk ice, for a wide range of conditions relevant to comets.

A number of three-dimensional Monte Carlo simulations were also run that considered the back-diffusion effect on {\em reaction rates} in a bulk-ice layer (as opposed to diffusion between layers), for reactions between like species (e.g. H+H). However, in this case the standard reaction rates were found to be consistent with the Monte Carlo simulations; back-diffusion reduced the reaction rates by a maximum factor around 1.5 for all except very thin ices (a few monolayers), for which the effect reached a factor of a few. For high occupation of reactants, the effect was removed entirely. These MC results, presented briefly in Appendix A, indicate that back-diffusion has only a weak effect on the rates of bulk-ice reactions driven by three-dimensional diffusive processes. Furthermore, the inclusion of alternative reaction partners (e.g. O, OH, etc.) would be expected to reduce the effect again, even in thin ices. 

Considering that the effects on rates are so small, we choose not to include any kind of parameterization of the bulk-ice back-diffusion effect on chemical reactions in the present models. For chemical reactions occurring in the surface layer of the comet, the back-diffusion correction of \citet{Willis2017} was implemented.

\subsubsection{Monte Carlo simulations of back diffusion between layers}\label{MCsims}

The ultimate goal of the MC simulations of bulk diffusion is to determine a parameter representing the average number of diffusion events (``hops'') undertaken by all diffusers of a particular chemical species, in a layer of chosen thickness, to produce the escape of {\em one} of those diffusers, either to the layer above or the layer below. We label this quantity the ``back-diffusion factor'', $\theta$, and it is involved in determining the rates of diffusive transfer of H and H$_2$ between layers in the comet model according to the expression
\begin{equation}
    \label{mod_rswap}
    R_{\mathrm{diff},m}(i) = k_{\mathrm{diff}}(i) N_{m}(i)  / \theta,
\end{equation}
with quantities defined as in Eqs.~(\ref{rdiff}) and (\ref{kdiff}). These rates should be divided by 2 when implementing separate upward and downward diffusive escape rates.

The MC model is functionally similar to that described by \citet{Garrod2017}, except that it has now been extended to many diffusers (which requires the ice to have an explicitly defined lateral size), and that it allows diffusers to escape from both the top and bottom interfaces (in the case of dust-grain ices, escape from the lower interface was blocked). The interstitial sites in the ice that the diffusers may occupy are assumed to take a simple cubic arrangement. For most of the MC simulations, we limit the maximum thickness to 30 monolayers (i.e.~30 sites), for computational efficiency. A lateral width of $N_{\mathrm{lat}}=15$ (i.e.~15 sites wide) is used for both lateral dimensions, with periodic boundary conditions applied across the four lateral surfaces to simulate a much larger ice. No periodicity is included for the vertical dimension.

Diffusion is simulated for a range of ice thicknesses up to the pre-defined maximum of 30~ML. The model begins by creating an ice with a thickness of 1~ML, and depositing one diffuser in a randomly-chosen site. This diffuser is then allowed to hop (i.e.~diffuse), with an equal probability of choosing any of the available directions. The number of hops taken is recorded until that diffuser leaves the ice, by hopping out of either the top or bottom monolayer. The diffuser is then replaced in another randomly-chosen site, and the process is repeated 500,000 times to ensure a good statistical sample, from which the mean value is recorded. Then a second diffuser is added and the simulations are re-run, and so on. For simulations with multiple diffusers, the \textit{total} number of hops of all diffusers is recorded, and all diffusers are replaced randomly in the ice once a single diffuser leaves. If a diffuser attempts to hop into a site that is already occupied, a swapping event occurs in which the two particles exchange places. This is counted as two hops, one for each particle.

Once the maximum number of diffusers (i.e.~$N_{\mathrm{lat}}^{2}N_{\mathrm{th}}$) has been reached in a given thickness, the model adds another layer onto the ice, and starts over again at one diffuser. This way, the maximum amount of coverage space is properly sampled, and a relationship detailing the magnitude of the back-diffusion effect as a result of both thickness ($N_{\mathrm{th}}$) and number of diffusers ($N_d$) can be determined. In order to derive expressions to parameterize the data, we make use of the {\tt Eureqa} modeling engine \citep{Schmidt2009,Eureqa}. \S \ref{back_results} details the results of the simulations and their parameterization.

\subsection{Heat transfer and solar approach}
\label{heat_meth}

G19 presented detailed models of the cold-storage phase of Oort cloud comets. In this work, we go beyond this, by modeling the chemistry during several solar orbits subsequent to cold storage; this requires knowledge of, in particular, the temperature behavior of the comet as a function of time. In order to calculate this behavior, the orbits of a comet of interest must first be calculated, which requires knowledge of its orbital elements. We used the {\tt PyEphem} package \citep{2011ascl.soft12014R} to load these orbital elements from the IAU Minor Planet Center\footnote{https://www.minorplanetcenter.net/data}. The orbital positions of the comet of interest (in this case, comet Hale--Bopp) were then calculated for one solar orbit, using the most recently-known orbital elements, which were originally downloaded in January 2020. These orbital elements were then assumed to be unchanged for subsequent orbits of the comet in this model.

\begin{table*}
    \centering
    \begin{tabular}{|c|c|c|}
        \hline
        Parameter & Value \\ \hline
        Inclination & 88.9912$^{\circ}$ \\ \hline
        Longitude of Ascending Node & 283.3585$^{\circ}$ \\ \hline
        Argument of Perihelion & 130.6487$^{\circ}$ \\ \hline
        Semi-Major Axis & 180.7048 AU \\ \hline
        Mean Daily Motion & 0.0004057$^{\circ}$ day$^{-1}$ \\ \hline
        Eccentricity & 0.99492735 \\ \hline
        Mean Anomaly & 0.0000$^{\circ}$ from perihelion \\ \hline
        Epoch Date & 03/29.6285/1997 \\ \hline
        Equinox Year & 2000 \\ \hline
        Perihelion Distance & 0.917867899 AU \\ \hline
        Aphelion Distance & 360.4929504 AU \\ \hline
        Orbital Period & 2429 years \\
        \hline
    \end{tabular}
    \caption{Physical parameters used to model the orbit of the comet. All parameters obtained from the IAU Minor Planet Center database, or calculated from them.}
    \label{orbparams}
\end{table*}

Hale--Bopp has a semi-major axis of 181 AU, an inclination of 89 degrees, and an eccentricity of 0.99. This results in a perihelion distance of 0.92 AU, an aphelion distance of 360 AU, and an orbital period of 2429 years. Detailed values are included in Table \ref{orbparams}. With the orbits of the comet calculated, the chemical model requires that the heat transfer throughout the cometary ice be calculated, in order to obtain temperatures at each layer, for each time point. This was done following the method of \citet{Herman1985}, in which the heat diffusion equation is solved in a one-dimensional fashion. The heat diffusion equation is given by

\begin{equation}
    \label{heatdiff}
    \rho c(T) \frac{\partial T}{\partial t} = \nabla [\kappa \nabla T],
\end{equation}

where $\rho$ is the material density, $c(T)$ is the specific heat, $\kappa$ is the thermal conductivity, and $T$ is the temperature. In the heat diffusion simulations presented here, we assume that the comet is spherical and uniformly illuminated by the Sun. Boundary conditions at the surface and the interior edge of the comet nucleus must be defined in order to obtain a solution. At the surface, the boundary condition is defined by conservation of energy:

\begin{equation}
    \label{surfacebound}
    \frac{\partial T}{\partial r}|_{r=R} = \epsilon \sigma T^4 - 
    \frac{(1-A)S}{d_H^2} \left\langle cos \, \xi \right\rangle - F_{\mathrm{ISRF}}.
\end{equation}

$R$ is the radius of the comet, $\epsilon$ is the emissivity, $\sigma$ is the Stefan-Boltzmann constant, $T$ is the temperature of the cometary surface, $A$ is the surface albedo, $S$ is the solar constant, $d_H$ is the heliocentric distance (which changes in time), $\left\langle cos \, \xi \right\rangle$ is the average value of the local solar zenith angle, and $F_{\mathrm{ISRF}}$ is the mean intensity of the interstellar radiation field. The first term on the right-hand side of Eq.~\ref{surfacebound} corresponds to the radiative cooling from the surface of the comet, while the second and third terms correspond to the incoming solar and interstellar radiation, respectively. Note that, in this formulation, we ignore the cooling effects of sublimation from the cometary surface, as well as thermal effects of phase changes within the nucleus (although sublimation is still included in the chemical model itself. These effects are left for future studies. 

The boundary condition in the interior of the comet is defined thus:

\begin{equation}
    \label{intbound}
    \frac{\partial T}{\partial r}|_{r=0} = 0,
\end{equation}

indicating that the heat flux becomes 0 at the center of the comet. Table \ref{heatparams} shows the values of the parameters used in the heat diffusion simulations and their literature references.

\begin{table*}
    \centering
    \begin{tabular}{|c|c|c|}
        \hline
        Parameter & Value & Ref. \\ \hline
        $\epsilon$ & 0.5 & \citet{Whipple1976}  \\ \hline
        $A$ & 0.04 & \citet{Mason2001} \\ \hline
        $\left\langle cos \, \xi \right\rangle$ & 0.25 & Uniform illumination \\ \hline
        $F_{\mathrm{ISRF}}$ & $2.67 \times 10^{-2}$ erg s$^{-1}$ cm$^{-2}$ & \citet{1990IAUS..139...63M} \\ \hline
        $\rho$ & 1 g cm$^{-3}$ & \citet{Guilbert-Lepoutre2011} \\ \hline
        $c_{w}$ & $7.4 \times 10^{4} \, T + 9.0 \times 10^{5}$ erg g$^{-1}$ K$^{-1}$ & \cite{Herman1985} \\ \hline
        $c_d$ & $1.2 \times 10^{7}$ ergs g$^{-1}$ K$^{-1}$ & \citet{Guilbert-Lepoutre2011} \\ \hline
        $h$ & 0.1 & \citet{Guilbert-Lepoutre2011} \\ \hline
        $\kappa_w$ & $7.1 \times 10^{-3} \, T$ erg s$^{-1}$ cm$^{-1}$ K$^{-1}$ & \citet{Kouchi1994} \\ \hline
        $\kappa_d$ & $4.2 \times 10^{5}$ erg s$^{-1}$ cm$^{-1}$ K$^{-1}$ & \citet{Ellsworth1983} \\ \hline
        $\psi$ & 0.3 & \citet{Guilbert-Lepoutre2011} \\
        \hline
    \end{tabular}
    \caption{Physical parameters used in the heat diffusion model.}
    \label{heatparams}
\end{table*}

As noted in Eq. \ref{heatdiff}, $c(T)$ is a function of temperature. The specific heat is defined as

\begin{equation}
    \label{specheat}
    c(T) = X_{w} \, c_{w} + X_{d} \, c_{d},
\end{equation}
where $X_{w}$ and $X_{d}$ are the mass fraction of water (as a proxy for the total ice) and dust, respectively, and $c$ is the specific heat of each component. For these simulations, we assume that $X_{w}$~=~$X_{d}$~=~0.5, following \citet{Guilbert-Lepoutre2011}. The expressions for $c_w$ and $c_d$ are given in Table \ref{heatparams}.

Similarly, $\kappa$ is also a function of temperature. In the simulations presented here, it is defined as

\begin{equation}
    \label{kappa}
    \kappa = \phi h \kappa_s,
\end{equation}
following \citet{Guilbert-Lepoutre2011}. In Eq. \ref{kappa}, $h$ is the Hertz factor, which scales the conductivity to account for reduced contact area between grains within the solid matrix. We set $h$ to 0.1. $\kappa_s$ is the conductivity of the cometary ice matrix, which is calculated as

\begin{equation}
    \label{icekappa}
    \kappa_s = x_w\kappa_w + x_d \kappa_d,
\end{equation}
where $x_{w}$ and $x_{d}$ are the volume fractions of water and dust, respectively, and $\kappa_{w}$ and $\kappa_{d}$ are the conductivities of amorphous water and dust, respectively. Table \ref{heatparams} shows their values. Note that, in this formulation, we assume that all ice in the comet is amorphous in nature. This is likely to be the case when the comet is in cold storage, but there will be phase changes when solar approach begins. However, our chemical model does not currently have the functionality to account for these phase changes, so we save their study for future work.

Finally, $\phi$ is a correction factor, first developed by \citet{Russell1935}, which accounts for the effect of pores on the thermal conductivity of a solid matrix. Its formula is given by

\begin{equation}
    \label{russell}
    \phi = \frac{\psi^{2/3}f + (1 - \psi^{2/3})}{\psi - \psi^{2/3} + 1 - \psi^{2/3}(\psi^{1/3} - 1)f},
\end{equation}
where $\psi$ is the porosity \citep[assumed to be 0.3, following][] {Guilbert-Lepoutre2011}, and $f$ is the ratio of the conductivity of pores to the solid matrix, $\kappa_p/\kappa_s$. The conductivity of the pores is calculated by

\begin{equation}
    \label{porekappa}
    \kappa_p = 4r_p \epsilon \sigma T^3,
\end{equation}
where $r_p$ = 10$^{-4}$ cm (the mean radius of a pore) and $\epsilon$ = 0.9 \citep{Guilbert-Lepoutre2011}.

Once the boundary conditions and parameters are defined, the heat diffusion equation can be solved through time. Firstly, to do this the physical space over which to solve Eq. \ref{heatdiff} must be defined; we chose $R$ = 30 km \citep[Hale--Bopp;][]{Fernandez2000}. Within this range, specific depth points must be defined, as well as the initial temperatures at those depths. In our temperature calculations, the ice is divided into 250 depth points, such that the final depth point ends at 30 km. The surface is considered its own layer, with a thickness of 1~cm, while the layers after that are spaced logarithmically in depth. Once the depths are set, the initial temperature is defined at each point. For this model, we simply assume a uniform temperature of 10~K throughout the ice at the initial time point of the simulation, corresponding approximately to Oort cloud conditions. 

Other heating effects, such as that produced by the radioactive decay of species including $^{26}$Al, are not considered in the present model. Heating by radioactive decay is primarily a deep mantle effect, in that greater heating occurs deeper into the comet. Our chemical model primarily considers surface and outer mantle chemistry where radioactive heating is weakest. It is possible that the addition of such effects could cause a stratification of volatiles \citep[e.g.][]{Prialnik2008}. Such might be noticeable after significant mass loss, but would have a relatively minor effect on the present model, which ignores outbursts that could be a major mass-loss contributor.

Once the initial conditions have been defined, the orbital position is allowed to evolve in time according to the specified cometary orbit. The impinging flux at the surface changes according to the heliocentric distance $d_H$, and the change in temperature is first calculated at the surface. Derivatives $\partial T/\partial r$ are then calculated throughout the ice. Once these derivatives have been calculated, the second derivatives are computed, and the temperatures at each depth in the ice are integrated using Gear's method. Once the temperatures are calculated for a full orbit, and the calculations are done again, continuing from those input conditions. Here, we calculate 5 orbits using Hale--Bopp's orbital parameters. \S \ref{heat_results} presents the results from these simulations, and discusses their inclusion in the cometary ice chemistry model.

\section{Results}
\label{results}

Results are divided into three subsections, corresponding to the back-diffusion calculations that are incorporated into the comet chemistry model (\S \ref{back_results}), the heat transfer calculations that determine the time- and depth-dependent temperature of the comet during solar approach (\S \ref{heat_results}), and the results from the cold storage and solar approach comet chemical kinetics models (\S \ref{model_results}) that make use of the results of \S \ref{back_results} and \S \ref{heat_results}.

\subsection{Back-diffusion between layers}
\label{back_results}

To determine the behavior of the back-diffusion factor, $\theta$, the three-dimensional Monte Carlo model described in \S \ref{MCsims} was run. The model provides the average total number of hops, $N_{\mathrm{hops}}$, by all diffusers in the system, required for a single diffuser to leave the ice. 
Although $N_{\mathrm{hops}}$ is precisely the quantity that we wish to express with our final formulation for $\theta$, we use the former label exclusively for the data produced by the MC models. The number of diffusers in the ice, $N_d$, is also equivalent to the quantity $N_{m}(i)$ in Eq.~(\ref{mod_rswap}).

\begin{figure}[h!]
    \centering
    \includegraphics[width=0.5\textwidth]{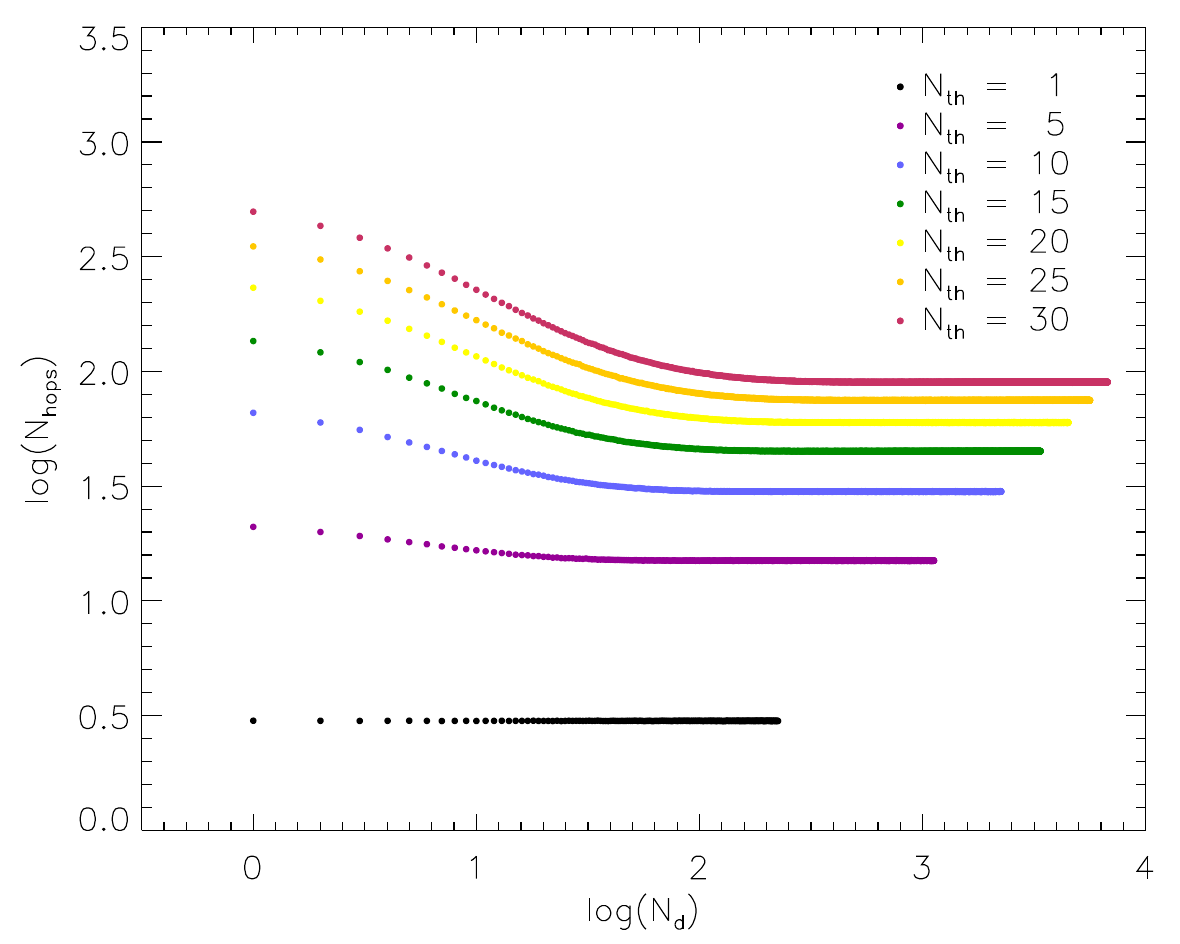}
    \caption{Raw data from 3D back-diffusion Monte Carlo code, using $N_{\mathrm{lat}}=15$. Data are color-coded according to thickness. The x-axis indicates the number of diffusers, while the y-axis displays the number of hops executed by all diffusers in the ice until one of the diffusers exits from the upper or lower layer of the ice.}
    \label{raw_bdiff}
\end{figure}

Figure \ref{raw_bdiff} shows the raw data for selected ice thicknesses, $N_{\mathrm{th}}$, between 1 and 30 ML. The intermediate thicknesses, although not shown, display similar behavior. The data range from $N_d = 1$ (a single diffuser) up to $N_d = N_{\mathrm{lat}}^{2}N_{\mathrm{th}}$ (full occupation).

Fig.~\ref{raw_bdiff} shows that, for ice thicknesses greater than one, $N_{\mathrm{hops}}$ falls with an increasing number of diffusers in the ice, $N_d$, until some lower limit is reached. This makes intuitive sense, as the more diffusers that are present in the ice, the greater the probability that some particles will randomly be placed initially close to the top or bottom interface of the ice. It may also be reasoned that, beyond some threshold in occupation, increasing the number of diffusers should no longer reduce the number of hops required; many of the additional diffusers would be sited deep within the ice, increasing the number of total hops without increasing the rate of successful exit from the interfaces.

It can also be seen from Fig.~\ref{raw_bdiff} that $N_{\mathrm{hops}}$ is constant for $N_{\mathrm{th}} = 1$, taking a value of 3.
For thicker ices, the number of hops required to escape is always greater than for thinner ices for any given value of $N_d$.

It is possible, without the need for a simulation, to derive the back-diffusion factor when the ice has the maximum number of diffusers (high occupation), meaning that back-diffusion will be at its minimum:
\begin{equation}
    \label{phihi}
    \theta_{\mathrm{hi}} = 3N_{\mathrm{th}},
\end{equation}
This expression is derived from the consideration that, if all sites are occupied, there is a total of $N_{\mathrm{lat}}^{2}N_{\mathrm{th}}$ diffusers that could diffuse, while only those in the upper or lower monolayers, a total of $2N_{\mathrm{lat}}^2$, could successfully diffuse out of the ice at either interface. Considering finally that one of these may only successfully exit with a probability of 1/6, Eq. \ref{phihi} becomes self-evident. The expression matches perfectly with the data in Fig.~\ref{raw_bdiff}, using the right-most point for each value of $N_{\mathrm{th}}$, corresponding to full occupation. The same expression gives a perfect match also to all of the data-points for the $N_{\mathrm{th}} = 1$ case, in which any upward or downward hop leads to the escape of the diffuser, meaning that back-diffusion is again minimized, regardless of the degree of occupation of sites.

\begin{figure}
    \centering
    \includegraphics[width=0.5\textwidth]{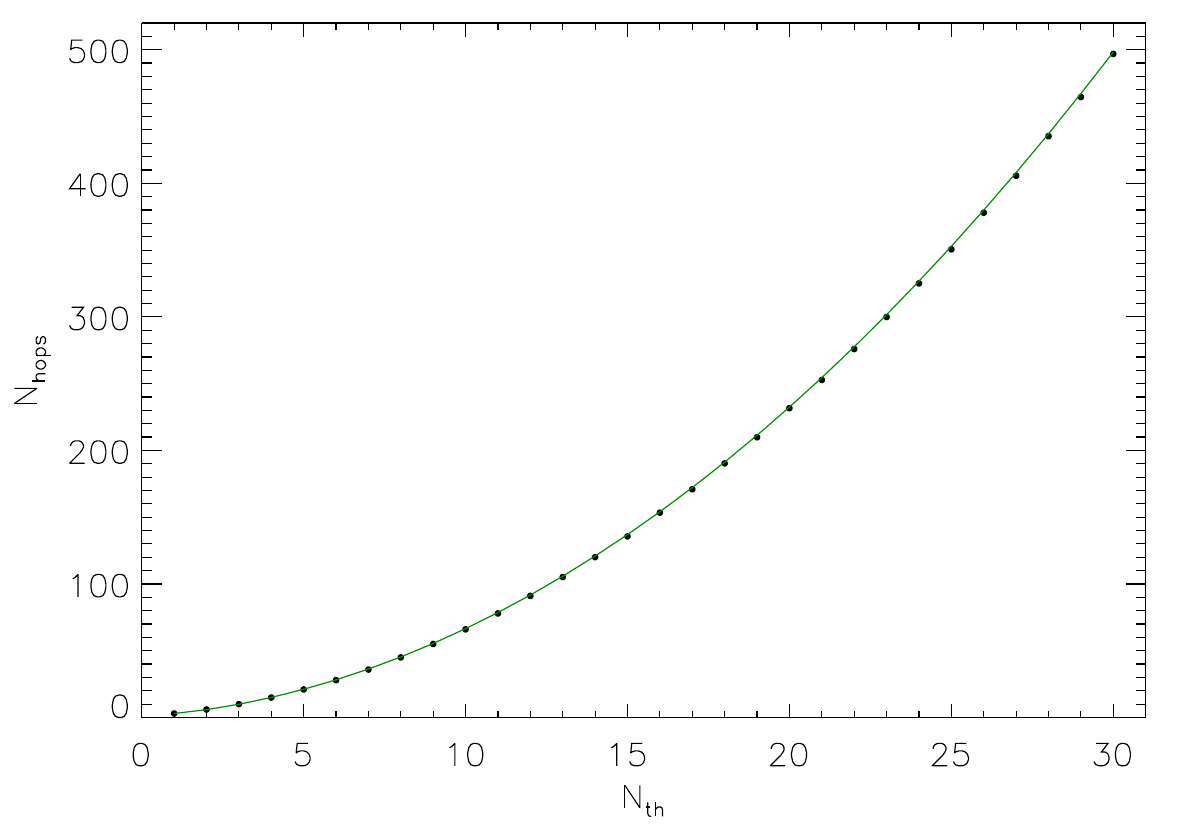}
    \caption{Monte Carlo data for single diffusers, using $N_{\mathrm{lat}}=15$, fit to Eq. \ref{philo}.}
    \label{onediff}
\end{figure}

At the other extreme, an expression for the single-diffuser case (low occupation) may be derived from the models. Based on the data partially shown in Fig.~\ref{raw_bdiff}, for the points $\log(N_{d})=0$, the single-diffuser back-diffusion factor relevant to the comet models is well fit by the expression
\begin{equation}
    \label{philo}
    \theta_{\mathrm{lo}} = 0.5N_{\mathrm{th}}^2 + 1.5846N_{\mathrm{th}} + 0.699,
\end{equation}
which is assigned a minimum value of 3, to ensure that the $N_{\mathrm{th}} = 1$ case is accurately reproduced. 
Fig. \ref{onediff} shows the fit for Eq. \ref{philo} with the full set of single-diffuser data ($N_{\mathrm{th}}= 1 - 30$).
Note that the single-diffuser expression shown in Eq.~(\ref{singlediff}), for a monolithic dust grain-surface ice mantle, as determined by \citet{Garrod2017}, corresponds to escape only from the upper layer. It is to be expected that the outcome in that scenario will be different from the case of two interfaces, when occupation is low and back-diffusion is significant.

\begin{figure}
    \centering
    \includegraphics[width=0.5\textwidth]{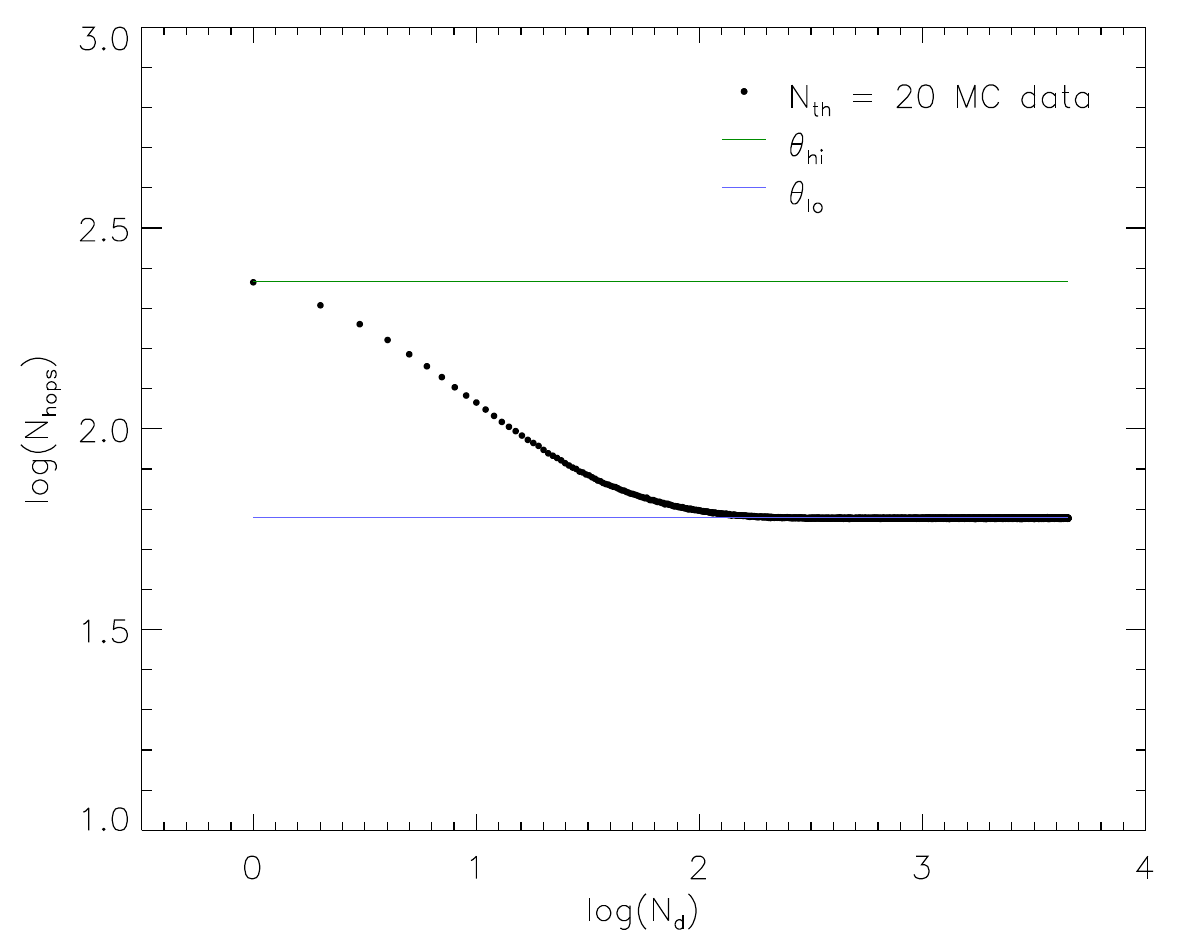}
    \caption{Data for back-diffusion simulations with $N_{\mathrm{th}}$ = 20, with maximum and minimum values obtained from Eq.~(\ref{phihi}) and Eq.~(\ref{philo}).}
    \label{transition}
\end{figure}

For larger values of $N_{\mathrm{th}}$, the behavior as a function of $N_d$ is more complicated. An example is shown in Fig.~\ref{transition}, for $N_{\mathrm{th}}$~=~20, where it is apparent that the data can be described by a function that converges to Eq.~(\ref{phihi}) at high numbers of diffusers and Eq.~(\ref{philo}) at low numbers of diffusers. 

In order to parameterize this relationship, we introduce the following simple function for the total back-diffusion factor, as a function of $\theta_{\mathrm{lo}}$ and $\theta_{\mathrm{hi}}$:
\begin{equation}
    \label{phitot}
    \log_{10}(\theta) = a\log_{10}(\theta_{\mathrm{lo}}) + (1-a)\log_{10}(\theta_{\mathrm{hi}}),
\end{equation}
where $a$ is a yet-to-be-defined switching function that describes the transition between the low- and high-occupation regimes. This function should depend on $N_d$ and $N_{\mathrm{th}}$.

We use the aforementioned {\tt Eureqa} modeling software \citep{Schmidt2009,Eureqa} to derive a relationship for $a$ from our MC model data. {\tt Eureqa} utilizes genetic programming to test several functional relationships between variables, attempting to find minima corresponding to the user's chosen error metric. For these data, we have chosen to minimize the weighted mean absolute error (WMAE). Most data-points are assigned a weight of 1. However, in order to ensure the proper behavior at both extremes (one diffuser, and maximum diffusers), we give a weight of 100 to each data point in our data set that corresponds to $N_d$~=~1 or $N_d$~=~$N_{\mathrm{lat}}^{2}N_{\mathrm{th}}$.

We also give a weight of 10 to those data points with $N_d$~=2 -- 10, as this parameter space is expected to be most important for simulations of cometary ice chemistry, based on test chemistry simulations. In addition, because we can theoretically calculate $\theta_{\mathrm{lo}}$ and $\theta_{\mathrm{hi}}$ for any thickness, we create dummy data corresponding to the single- and maximum-diffuser case for each ice thickness up to $N_{\mathrm{th}}$=10,000.

Using these input data, {\tt Eureqa} provides the following function for $a$,
\begin{equation}
    \label{switch}
    a = N_d^{-0.418\sqrt{\frac{N_d}{N_{\mathrm{th}}}}}, \nonumber
\end{equation}
which achieves a WMAE of $\sim$6e-5. With this basic relationship established, we conducted a further parameter search using the expressions
\begin{equation}
    \label{switch1}
    a = \left(N_d\right)^{b}
\end{equation}
\begin{equation}
    \label{switch2}
    b = -f_{1} \left( \frac{N_d}{N_{\mathrm{th}}} \right)^{f_{2}}.
\end{equation}

\begin{figure}[h!]
    \centering
    \includegraphics[width=0.5\textwidth]{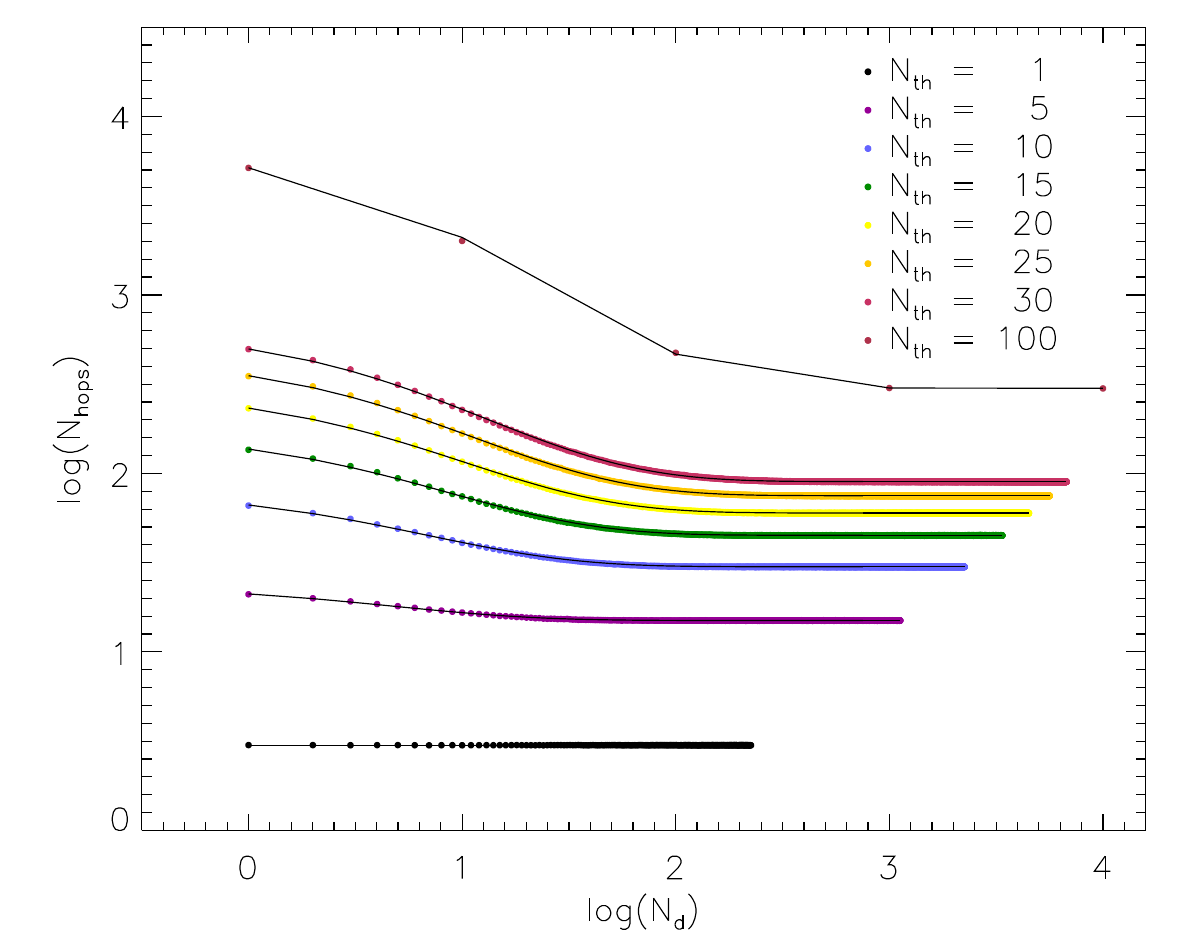}
    \caption{Data for back-diffusion simulations up to $N_{th}$~=~100; black lines indicate the fitted values, $\theta$.}
    \label{nth100}
\end{figure}
Values $f_{1} = 0.4051$ and $f_{2} = 0.390$ were found to provide the best match, attaining a maximum fractional error per data point of $\sim$2\% for the entire dataset, with an average error of $\sim$0.2\% for each $N_{\mathrm{th}}$ value. Note also that a certain degree of computational error exists in the MC data, meaning that a perfect match to all datapoints is not achievable.

Eqs.~(\ref{switch1}) and (\ref{switch2}) can be inserted into Eq.~(\ref{phitot}) to calculate the back-diffusion factor for any arbitrary $N_d$ and $N_{\mathrm{th}}$ values. 

Figure \ref{nth100} displays the Monte Carlo data along with the curves produced by the fit. In addition to the $N_{\mathrm{th}}$ = 1 -- 30 data, we also ran the Monte Carlo diffusion model with $N_{\mathrm{th}}$~=~100, to test the relationship at greater thicknesses, also shown in Figure \ref{nth100}. Due to computational costs, the  $N_{\mathrm{th}}$~=~100 model was run for five values of $N_d$, ranging logarithmically from 1 to $10^4$. An excellent match is produced in this case also.

To ensure no dependence on the chosen lateral size of the ice used in the MC simulations, $N_{\mathrm{lat}}$, we also ran the models for the $N_{\mathrm{lat}}$ = 8 and 10 cases. The model fit obtained from the $N_{\mathrm{lat}}$ = 15 data works equally well for the other two cases, demonstrating that the lateral size of the periodically bounded ice is not a significant parameter in the back-diffusion.

Incorporation of Eqs.~(\ref{mod_rswap}) and (\ref{phihi}) -- (\ref{switch2}) into the rate-equation treatment for diffusion between layers in the cometary ice model is simple. Abundances of species within the ice layers in the comet model are given in units of ML. These abundances must first be converted into a raw number of diffusers, $N_d$, in order to calculate the value of $\theta$. To do this, the volume of 1 ML of ice must be calculated. Here we make a simple approximation, by calculating the surface area of the comet given its radius $R$ \citep[30~km for Hale--Bopp;][]{Fernandez2000} and multiplying that by the thickness of 1 ML. In our treatment, 1 ML is equivalent to $3.215 \times 10^{-10}$~m, given that cometary ice may be well represented by amorphous solid water ice \citep{Brown1996}. In reality, the comet will include various phases of water ice, and would be porous in nature; we omit these considerations in this simple calculation. Once the volume of a monolayer is determined, it is divided by the volume of one \ce{H2O} molecule to obtain the number of molecules in one ML of ice.
From this, $N_{d}$ for the diffuser of interest (H or \ce{H2}) can be calculated. 
The same modification is made for the calculations of the diffusion rates between all layers in the cometary ice.

Complementing the results presented above, we conducted similar MC models and analysis to parameterize $\theta$ for all values of $N_d$ and $N_{\mathrm{th}}$ for the case of interstellar dust grain ices, using a monolithic ice mantle with only an upper interface. Parameters $f_{1} = 0.3126$ and $f_{2} = 0.390$ were obtained in that case; this new treatment should supercede that presented by \cite{Garrod2017} for use in interstellar astrochemical models.

\begin{figure*}
    \centering
    \includegraphics[width=0.49\textwidth]{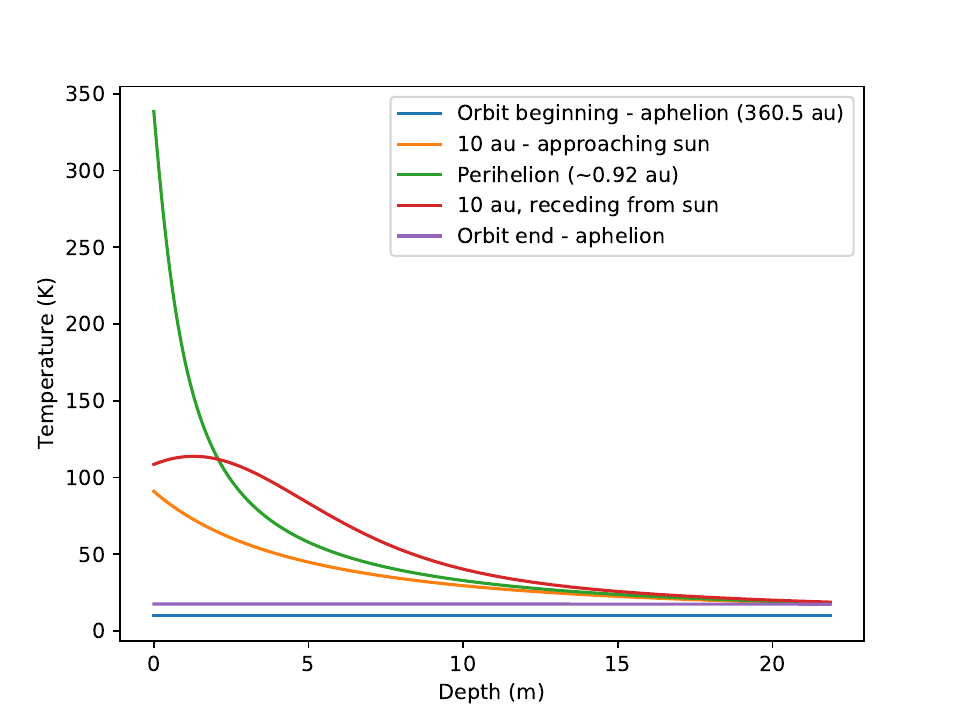}
    \includegraphics[width=0.49\textwidth]{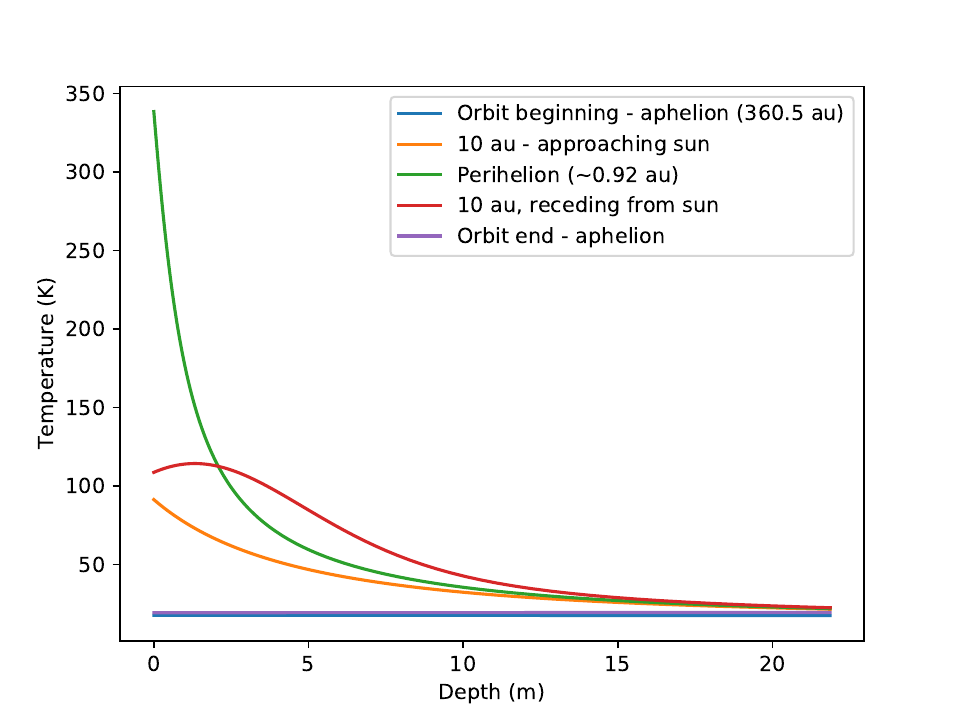}
    \caption{Temperature through the first $\sim$20~m of ice for the first (left panel) and second (right panel) orbital evolution of Hale--Bopp, based on the model of \S \ref{heat_meth}. The plots are color-coded for each orbital position. Subsequent orbits are very similar in behavior to the second orbit.}
    \label{orbit1}
\end{figure*}

\subsection{Heat transfer simulations}
\label{heat_results}

The method outlined in \S \ref{heat_meth} was used to solve the temperature evolution of Hale--Bopp ($R$~=~30 km) for 5 orbits. Fig.~\ref{orbit1} (left panel) shows the results for the first orbit, at five different radial distances from the sun. The maximum temperature for this first orbit, attained at perihelion, is $\sim$338~K, achieved in the surface layer. As the comet recedes from the sun, it cools from the surface first, leading to peak temperatures that occur a few meters into the ice. By the time the orbit is complete, the temperature is mostly uniform throughout the first 20~m, slightly elevated from the initial value of 10~K, at a temperature of $\sim$18~K. Temperatures greater than 15~K are maintained to a depth of $\sim$88 m (not shown here, in order to concentrate on the heating at the surface). The shapes of these temperature profiles are qualitatively similar to the one-dimensional calculations of \citet{Herman1987} (see their Fig.~2), although their methods were slightly more complicated, as they included a crystalline component to their ice, as well as sublimation in their version of Eq.~\ref{surfacebound}. 

The remaining four orbits look qualitatively similar to Figure \ref{orbit1}, and are very similar to each other; the right panel of Figure \ref{orbit1} shows the temperature behavior during the second orbit. The maximum temperature achieved at perihelion is $\sim$338~K for each orbit. One difference between orbits is that the temperature evolves slightly further down into the cometary ice for each evolution, though this effect is small. For example, the temperature at a depth of 20 m is 18~K at the end of the first orbit, whereas at the end of the fifth orbit it is $\sim$21~K. Thus there is some small hysteresis in the ice temperature.

The time- and depth-dependent results of these simulations are read into the model of cometary ice chemistry at run time, to be converted into temperatures representative of the 25 chemically distinct ice layers.
The layer widths in the heat transfer simulations are not the same as those in the chemical model.
The first 16 layers in the comet chemistry model all fall within the 1~cm-thick surface layer of the heat-transfer model; thus those first 16 layers all have the same temperature throughout the evolution. The deeper, thicker layers in the chemical model sometimes intersect multiple depths whose temperatures are calculated independently in the heat-transfer model. In these cases, the temperature used for each chemical-model layer is the depth-weighted average of the calculated values from the heat-transfer model.

\subsection{Results of updated chemical models}
\label{model_results}

The basic chemical model of G19, outlined in \S \ref{basic_meth} and \S \ref{nonthermal_meth}, was updated to include the back-diffusion treatment of \S \ref{back_results}, along with updates to the chemical network and the inclusion of non-thermal chemical mechanisms. These additions and alterations to the model produce significant variations in the chemical behavior of the cold storage phase, compared with G19, as described below in \S \ref{cs_results}. The chemical behavior during the new dynamical phase, occurring after the cold storage and using temperature profiles obtained in \S \ref{heat_results}, is described in \S \ref{sa_results}.

\subsubsection{Cold storage phase}
\label{cs_results}

\begin{figure*}
    \centering
    \includegraphics[width=0.325\textwidth]{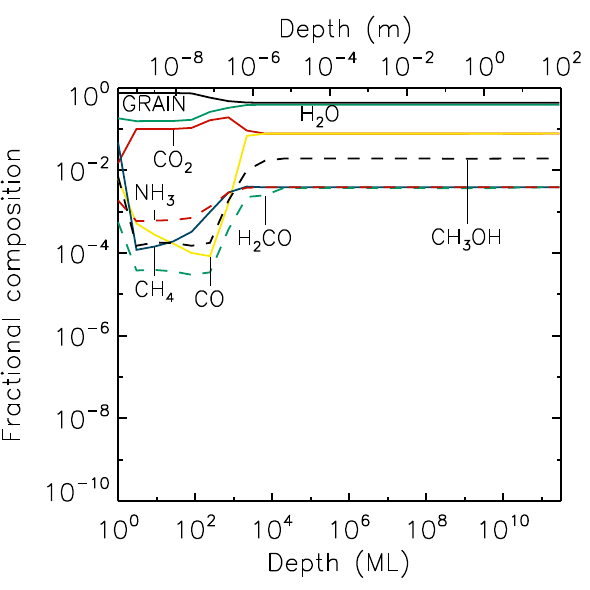}
    \includegraphics[width=0.325\textwidth]{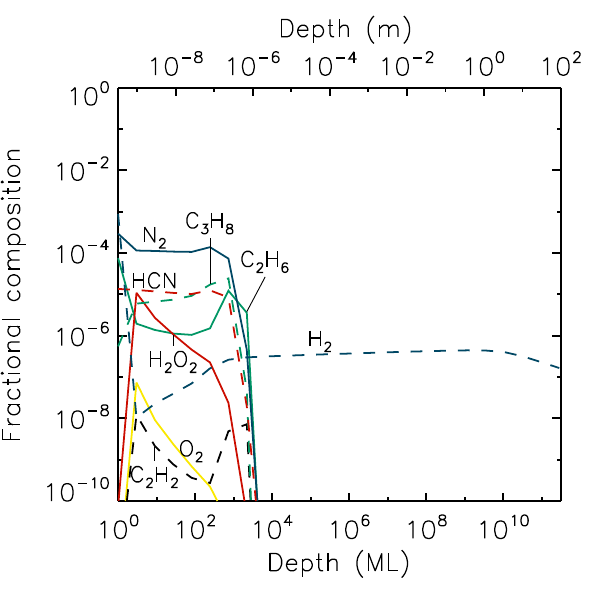}
    \includegraphics[width=0.325\textwidth]{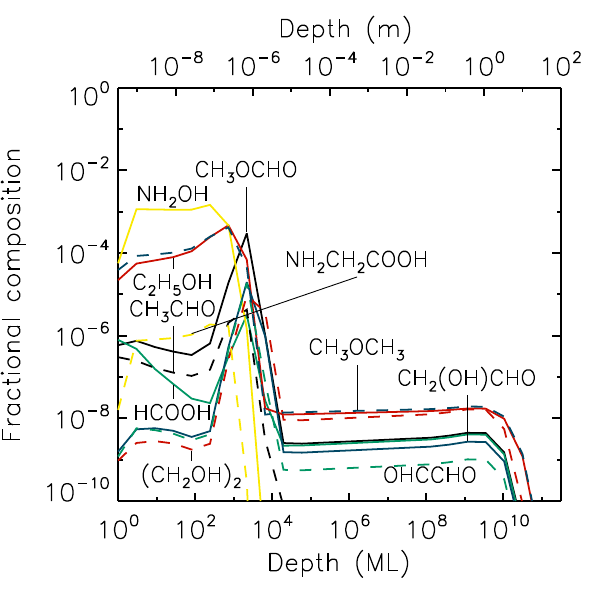}
    \caption{Fractional abundances of selected ice species at \(10^{6}\) years into the cold storage phase of the updated model, plotted with respect to depth into the comet surface; the surface (outermost) layer is at the left in each panel. The x-axes show the depth into the comet in units of monolayers (ML), shown on the bottom, and meters (m), shown on the top. The left panel shows species that are initially present in the ice, while species in the middle and right panels are products of the chemistry during cold storage.}
    \label{hb1e6}
\end{figure*}

\begin{figure*}
    \centering
    \includegraphics[width=0.325\textwidth]{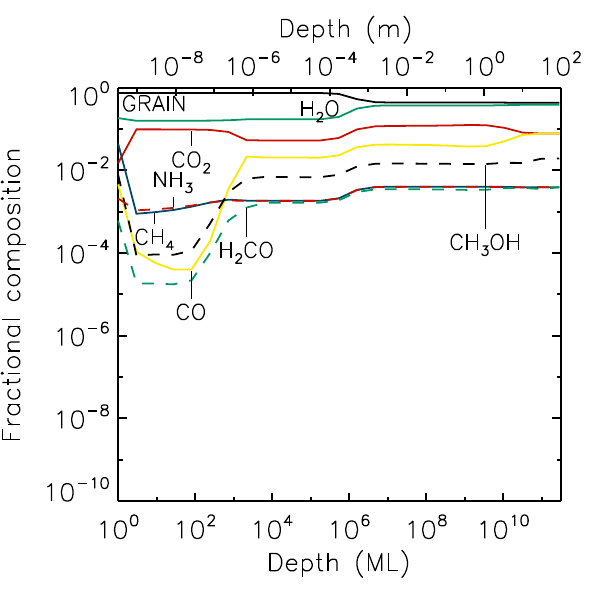}
    \includegraphics[width=0.325\textwidth]{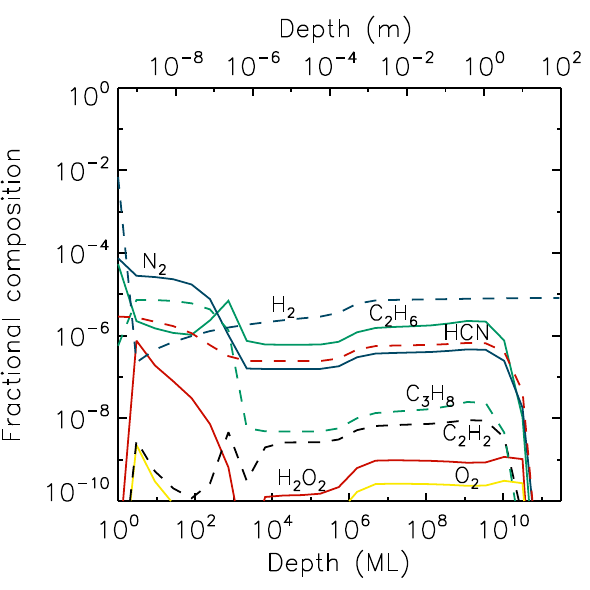}
    \includegraphics[width=0.325\textwidth]{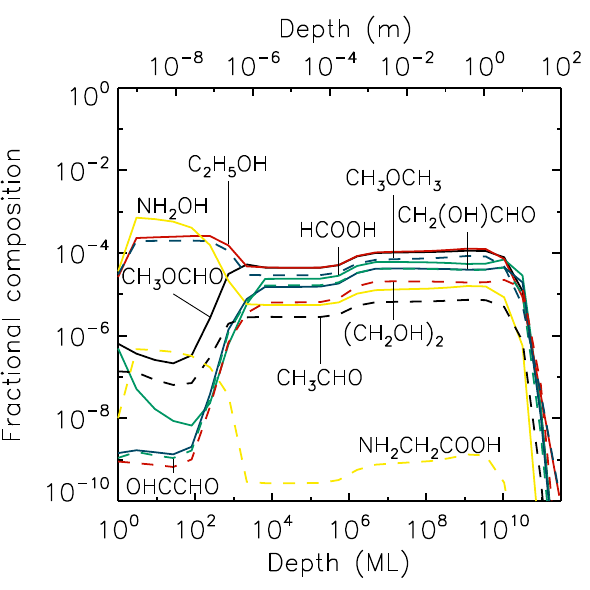}
    \caption{Fractional abundances of selected ice species at \(4.5 \times 10^{9}\) years into the cold storage phase of the updated model, plotted with respect to depth into the comet surface; the surface (outermost) layer is at the left in each panel. The x-axes show the depth into the comet in units of monolayers (ML), shown on the bottom, and meters (m), shown on the top.}
    \label{hb4e9}
\end{figure*}

\begin{figure*}
    \centering
    \includegraphics[width=0.325\textwidth]{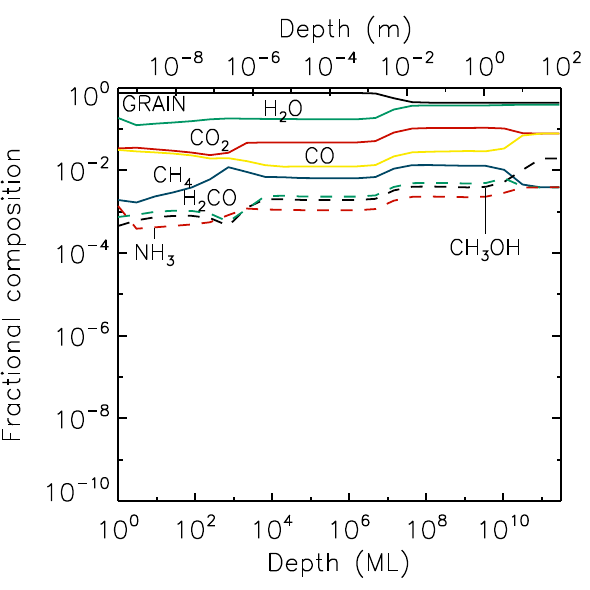}
    \includegraphics[width=0.325\textwidth]{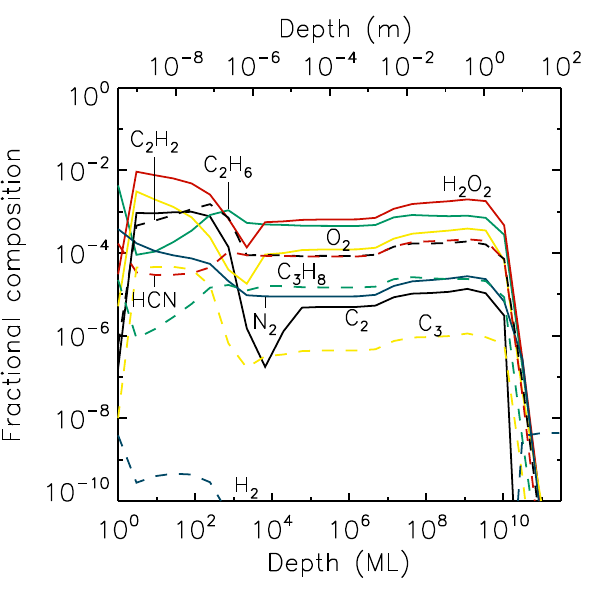}
    \includegraphics[width=0.325\textwidth]{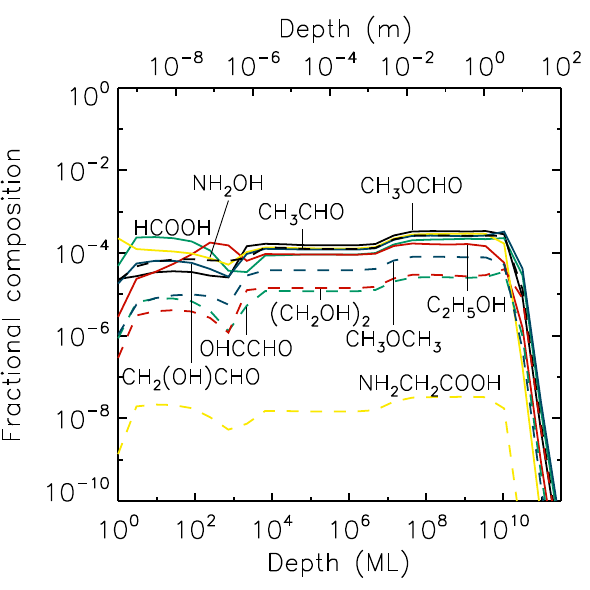}
    \caption{Fractional abundances of selected ice species at \(4.5 \times 10^{9}\) years into the old G19 simulation, for comparison with the new data in Figure \ref{hb4e9}.}
    \label{G19-fig}
\end{figure*}

The previous chemical model presented by G19, which was compared with data for comet Hale--Bopp, had a different endpoint than the updated version presented here, running to 5~Gyr as opposed to the 4.5~Gyr in the updated model. The latter value was chosen to provide a slightly more accurate value for the age of the Solar System. Thus, data from the G19 model was re-plotted at 4.5~Gyr so as to compare directly with the new model endpoint for the cold storage phase. The plots in our Figure \ref{hb1e6} show results at $10^6$ yr into the cold storage phase, and may be compared with those of G19, in their Figure 12, panels d--f. Figure \ref{hb4e9} shows abundances at the end-time of the model, with Figure \ref{G19-fig} showing the re-plotted data at the equivalent time in the G19 model. Species shown in the left panels of Figs.~\ref{hb1e6}--\ref{G19-fig} are those that are present in the ice at the beginning of the models. Species in the middle panels are exclusively products; the right panels show some of the more complex product species.

At 10$^6$ years into the cold storage phase (Fig.~\ref{hb1e6}), the upper 1~$\mu$m of material shows the most variation in chemical composition, as was the case in the G19 models. Aside from in the very surface layer, most of the species that are initially present in the ice (left panel) see a drop in their fractional abundances by this time. The only exception to this is \ce{CO2}, which is instead enhanced at the expense of \ce{H2O} and \ce{CO}; photodissociation of water produces an OH radical that may immediately react with nearby CO in the ice (via the PDI mechanism). It is also notable that while \ce{H2O} fractional abundance drops in the upper micron, it remains the dominant molecule at all depths and times. The fraction of dust grains in the upper micron is also increased, although it only reaches its maximum allowed value ($\sim$74\%) in the upper 10$^{-8}$~m or so. The lower abundance of \ce{H2O} and other species in the upper layers is partly due to this crowding out by dust grains. The increased dust concentration is ultimately caused by the destruction and/or sublimation of volatile species at the surface, while the dust grains are left in place.

The production of new species in the upper $\sim$1~$\mu$m is dominated by photodissociation by interstellar UV; chemistry in the deeper layers is dominated by the dissociative action of cosmic rays, which have a substantial effect to depths on the order of around 10~m.

The product-species \ce{H2O2}, \ce{N2}, \ce{HCN}, and some hydrocarbons reach fractional abundances of around 10$^{-6}$--10$^{-4}$ down to $\sim$1~$\mu$m depth. Much of the hydrogen peroxide is formed through photodissociation-induced (PDI) association reactions between OH radicals, while molecular nitrogen is formed through a similar reaction between NH radicals, which are the product of ammonia photodissociation. Dissociation of CO produces carbon atoms that may react with the dissociation products of ammonia, ultimately forming HNC and HCN. Atomic oxygen and OH may combine in the ice to form O$_2$H, which can be photodissociated or react with atomic H to produce O$_2$. Hydrocarbons of various hydrogenation states may be formed via the recombination of the dissociation products of methane (CH$_4$), while diffusive atomic H may also hydrogenate these species or abstract a hydrogen atom from them.

Some COMs, such as ethanol (\ce{C2H5OH}) and dimethyl ether (\ce{CH3OCH3}) also reach similar abundances down to $\sim$1~$\mu$m depth; however, they also maintain fractional abundances of up to \(10^{-8}\) down to depths greater than 1~m or so. In the upper micron, these species are primarily formed via PDI reactions, and they reach something close to their peak abundances in less than 1000 yr. Ethanol and dimethyl ether in particular may build up their abundances by photodissociation of methanol to produce one of the radicals CH$_2$OH or CH$_3$O, which can combine with a methyl (CH$_3$) radical originating from photodissociation of methane or methanol. Photodissociation of methane may also produce the diradical methylene (CH$_2$), which can react directly with methanol to form dimethyl ether or ethanol. In the deeper layers, dissociation caused by cosmic rays results in similar production routes.

Other COMs, such as ethylene glycol (\ce{(CH2OH)2}) and glycolaldehyde (\ce{CH2(OH)CHO}) maintain abundances below \(10^{-6}\) before abruptly spiking up by roughly two orders of magnitude at a depth of 1~$\mu$m and then immediately falling to abundances of \(10^{-8}\) similar to the COMs previously mentioned. These species are primarily formed via three-body (3-B) reactions, i.e. the mechanism by which the product of a preceding reaction reacts immediately with some other nearby species. More specifically, mobile atomic H may react with methanol (CH$_3$OH) in the ice, abstracting a hydrogen atom to produce the CH$_2$OH radical, which can immediately react with some other nearby CH$_2$OH to form ethylene glycol. Although direct photodissociation of methanol to produce CH$_2$OH does occur, the H-abstraction reaction provides more CH$_2$OH, due to the high rate of H production as the result of water photodissociation.

Repetitive H-abstraction from ethylene glycol by other H atoms can lead directly to glycolaldehyde and glyoxal (OHCCHO); ethylene glycol production is the main route into the formation of all three of these molecules in the upper 1~$\mu$m. In the deeper layers, 3-B radical-recombination reactions produce each of the three molecules directly, i.e. through the addition of HCO radicals to produce glyoxal, the addition of HCO to CH$_2$OH to produce glycolaldehyde, or CH$_2$OH radicals combining to form ethylene glycol. HCO itself is mostly formed by the reaction of mobile H with CO, while CH$_2$OH is again mainly produced by H-abstraction from methanol by H atoms.

The spike effect in certain COM abundances around the 1~$\mu$m mark is caused by the competing formation and destruction produced by interstellar UV photons, which penetrate only to this limited depth. The photons enhance the production of COMs by producing radicals from simpler species, which then react together; but the UV also directly dissociates the product COMs, especially in the layers closest to the surface, where there is less absorption by the dust. The H atoms that are the product of photodissociation also abstract H from existing COMs, resulting in further destruction and conversion to other products. In this way, while the overall production of COMs like methyl formate (CH$_3$OCHO) via radical addition falls with depth into the ice, the coincident drop in H production from all sources results in yet slower destruction, pushing up the abundance toward the 1~$\mu$m threshold. Beyond this point, the production of reactive radicals and of destructive atomic H relies solely on GCR-induced dissociation. Over time, the much lower rates involved with this mechanism gradually allow COMs to grow to substantial abundances in the deeper layers.

For some species, abundances drop dramatically in the uppermost (i.e. surface) layer, as compared with the layers directly beneath, or in some cases may instead be enhanced in that surface layer. This effect is caused by a combination of the ability of surface species to freely diffuse (with rates dependent on the temperature), and thus react, while volatile species may also desorb entirely from the surface layer, which is also not possible for species in the deeper layers.

While a number of more chemically complex species begin to reach substantial abundances to depths $>$1~m by \(10^{6}\) yr, as the result of cosmic ray-induced chemistry, some of the more intermediate-size ice species remain at low abundances at depths greater than one micron. However, by the end of the model run most species have reached more substantial abundances at these depths.

It is worth noting that the fractional molecular hydrogen abundance is relatively stable right through the ice, which is due to its ability to diffuse, allowing it to reach the very deepest layers. The H$_2$ abundance at all depths increases further by the end-time of 4.5 Gyr.

As may be expected, by 4.5~Gyr the chemical enhancement throughout the ice is much more far-reaching than at 1~Myr, as shown in Figure \ref{hb4e9}; the longer timescale allows the slow cosmic ray-driven chemistry to convert more of the simple species into COMs, while also enhancing some other species that are formed by the destruction of those COMs themselves. Although in general the fractional abundances of ice species within the first micron have decreased somewhat compared to Figure \ref{hb1e6}, some species such as ethane and propane do not experience much change at all in the upper layers, instead only exhibiting differences in the deeper layers. In those deeper layers, below the first micron, the abundances of most product species shown in the middle and right-hand panels in Fig.~\ref{hb4e9} have increased by the end of the cold storage phase. 

It may be noted also that the concentration of dust-grain material extends yet deeper at the end of the model run, achieving its maximum allowed value down to as deep as 100~$\mu$m. In the G19 model, shown in Figure \ref{G19-fig}, this enhancement reaches deeper, to around 1000 $\mu$m, due to the different diffusion behavior of volatiles combined with the weaker overall dissociation rates (due to recombination between heavy photoproducts). In both models, a slight depression may be observed in the abundances of other species down to this depth, caused again by the crowding effect of the dust grains (rather than any explicit chemical effect). The further concentration of dust abundances in the upper layers by the end of the model also induces a slight shift in the reach of interstellar UV to somewhat shallower layers (compared with earlier times), as the photons are less able to penetrate, leading to lower dissociation rates and a less active photon-induced chemistry. Again, while the larger COMs show a closer correspondence between the new model and the G19 model, intermediate-sized hydrocarbon species, and some others, are generally lower in the new models. Interestingly, in the new model, glycine (NH$_2$CH$_2$COOH) shows a somewhat stronger production in the upper micron than was found in the G19 model, while at greater depths it is rather weaker than in G19; the G19 model produced glycine at a fairly stable value of around 10$^{-8}$ at all depths. As noted by G19, even those values are insufficient to explain the presence of glycine in cometary ices, indicating that the glycine in e.g. 67P likely has a more primordial origin, rather than it being formed in situ during cold storage.

Comparing the old and new models more broadly, the most notable differences can be found in the upper micron. Below these depths, the general order-of-magnitude behavior remains the same between the models for simple species (left panels in Figs.~\ref{hb1e6}--\ref{G19-fig}) at \(10^{6}\) years, and for both simple and complex molecules (right panels), as well as some intermediate-size species (middle panels) at \(4.5x10^{9}\) years. Additionally, while in G19, \ce{H2} had mostly been depleted in the bulk ice by 4.5 Gyrs, the updated model maintains a steady level and even increased fractionally from the previous milestone at \(10^{6}\) years.

Within the upper micron of the ice, most species have decreased in fractional abundance compared to the results in G19. Notable exceptions to this are \ce{CO2}, \ce{H2}, \ce{N2}, \ce{NH2OH}, and \ce{NH2CH2COOH}. 
Meanwhile, \ce{C2} and \ce{C3} abundances are significantly lower at all depths and fall below the lowest shown fractional abundances. As a result, \ce{C2H4} and \ce{C3H6} were chosen as substitutes for illustrative purposes. At \(10^{6}\) years there is a spike in fractional abundance right at the one micron barrier while maintaining similar abundance values on either side of the peak, while others simply decrease in abundance below the one micron threshold.

An important divergence between the new models and the G19 models at 10~K is that most molecules produced in the upper micron are unable to diffuse deeper into the ice in the new treatment, which now only allows bulk diffusion for H and \ce{H2}. Although molecular diffusion on internal pore surfaces, which is a different process, may be plausible in the case of cometary ices, such behavior is beyond the scope of the present model. The G19 treatment assumed that bulk diffusion in the ice was a swapping process, which is now deprecated. The reasons for limiting bulk diffusion in this way are explained in more detail by \cite{G22}, for the case of interstellar ices.

Another difference in the chemical treatment between the new and old models, noted in \S~2.2, involves the outcome of photodissociation events, in the case that one or other of the photoproducts does not immediately react with some other species already present in the ice via the nondiffusive PDI mechanism. In such cases, the new treatment allows these (trapped) photoproducts immediately to recombine (except in cases where H or \ce{H2} are produced, as these are free to diffuse within the bulk ice). The same change is made also for the equivalent radiolysis-induced dissociation that has greater influence in the deeper layers. The change results in a rather lower rate of radical production in the ice overall. Consequently, a number of the intermediate-size radicals shown in the middle panels of Figure~\ref{hb1e6} (and Figure~\ref{hb4e9}) achieve different abundances than in the G19 models. A specific example is the photodissociation of methanol (CH$_3$OH) in the upper layers; while the ejection of an H atom from methanol may produce a free CH$_3$O or CH$_2$OH radical in the ice, the branch that produces CH$_3$ + OH will in most cases result in recombination of these two radicals, as the chances of some other reactive radical being available nearby are much less than unity (although generally on the order of 1\% or less). Methanol therefore serves as a much weaker source of the CH$_3$ radical than in the G19 models. This reduces the production of ethane (C$_2$H$_6$) in the ice, as well as other simple hydrocarbons to which it is chemically related, when comparing with the G19 results. The larger COM species, such as ethylene glycol, (CH$_2$OH)$_2$, are formed from methanol radicals such as CH$_2$OH, so their production is less strongly affected by the change.

\begin{figure*}
    \centering
    \includegraphics[width=0.325\textwidth]{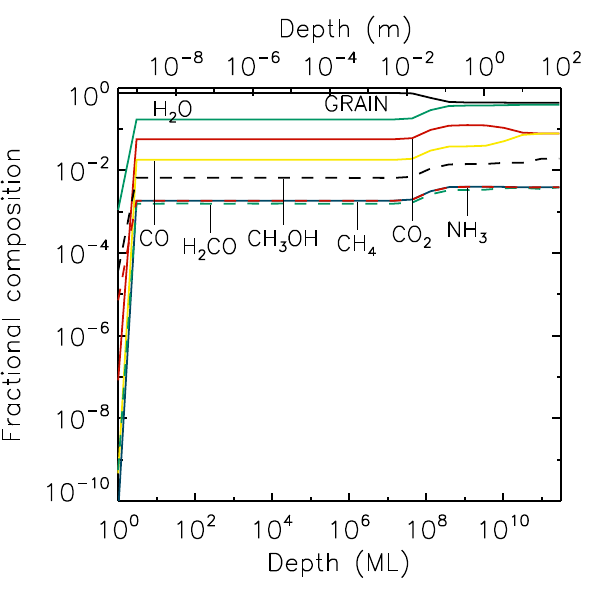}
    \includegraphics[width=0.325\textwidth]{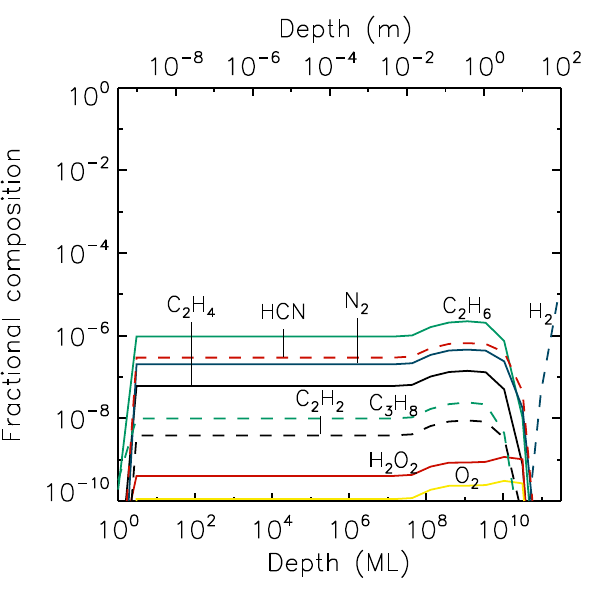}
    \includegraphics[width=0.325\textwidth]{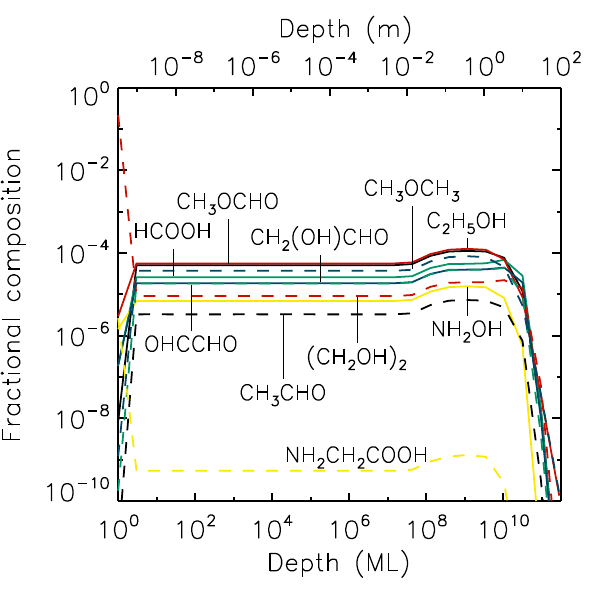}
    \\[\smallskipamount]
    \includegraphics[width=0.325\textwidth]{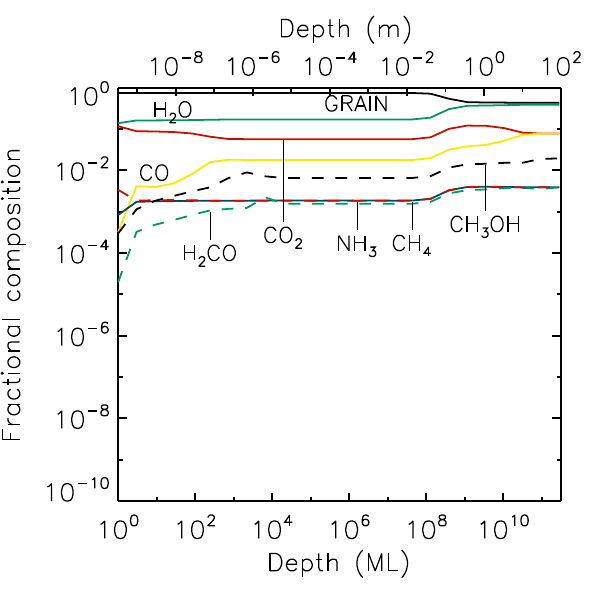}
    \includegraphics[width=0.325\textwidth]{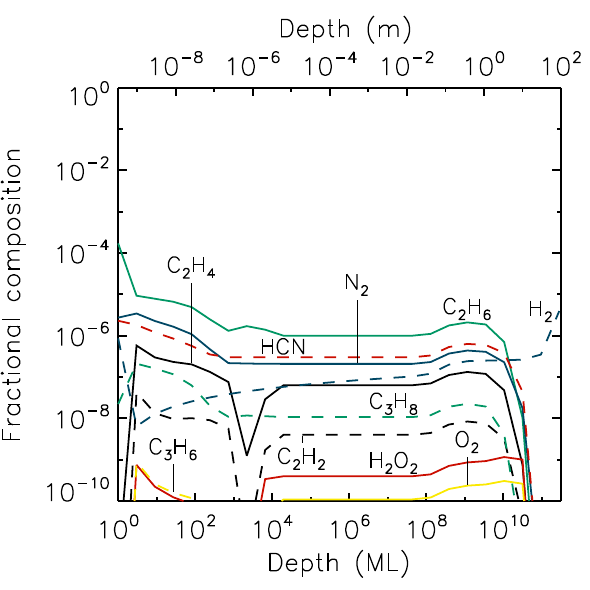}
    \includegraphics[width=0.325\textwidth]{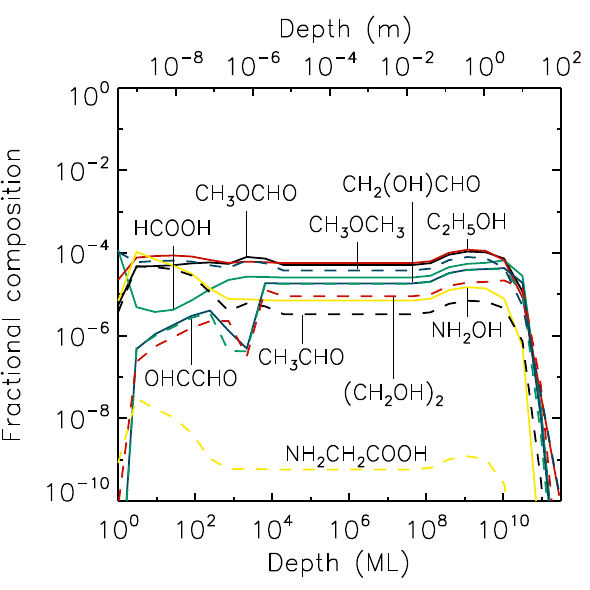}
    \caption{Fractional abundances of select ice species for the {\it first} of five solar approaches; Top: Abundances at first perihelion. Bottom: Abundances at first aphelion after starting point. The y-axis of each plot corresponds with the ice surface. 
    The x-axis is the depth, from the surface, in terms of monolayers (ML) shown on the bottom and meters (m) shown on the top. Each column represents a different subset of species grouped based on size and relevance.}
    \label{hbSA1}
\end{figure*}

\subsubsection{Solar approach phase}
\label{sa_results}

\begin{figure*}
    \centering
    \includegraphics[width=0.325\textwidth]{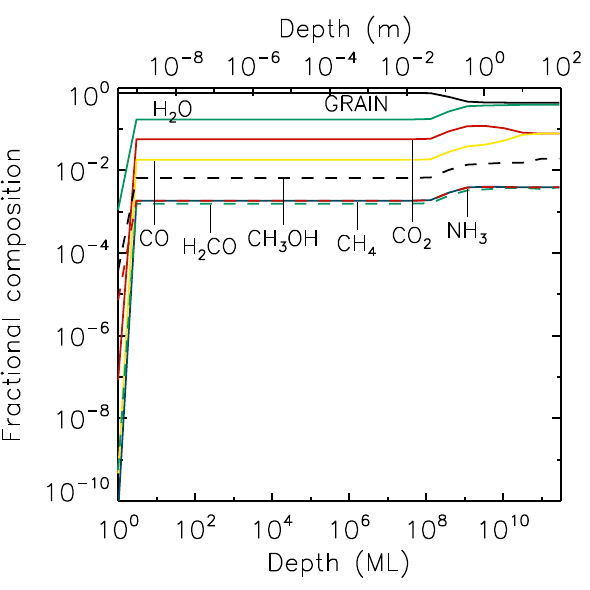}
    \includegraphics[width=0.325\textwidth]{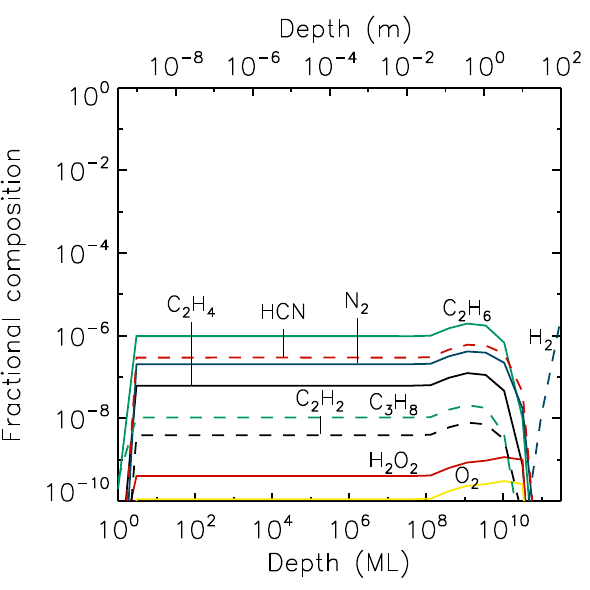}
    \includegraphics[width=0.325\textwidth]{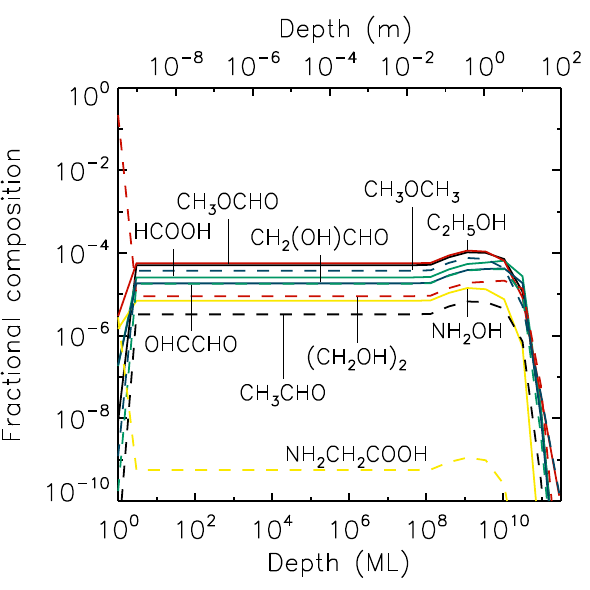}
    \\[\smallskipamount]
    \includegraphics[width=0.325\textwidth]{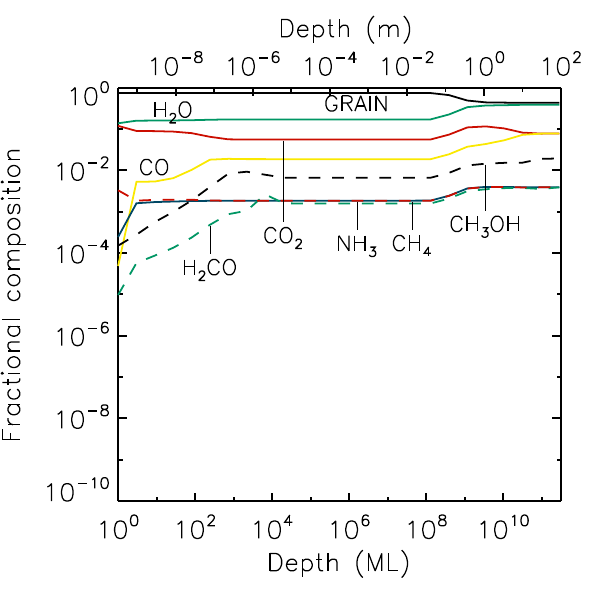}
    \includegraphics[width=0.325\textwidth]{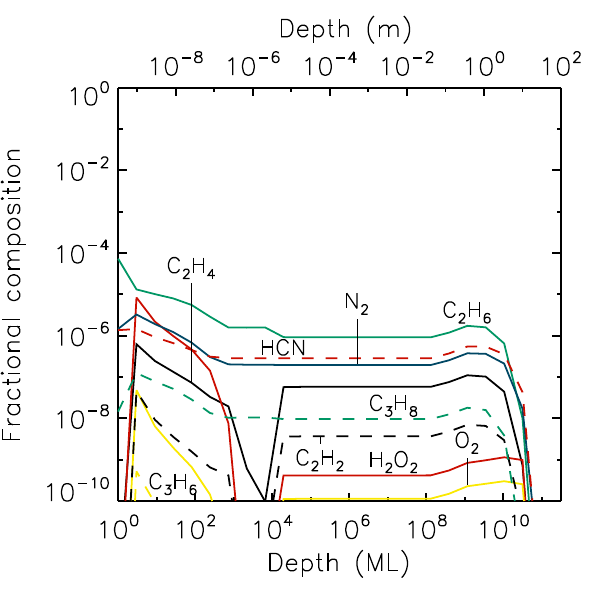}
    \includegraphics[width=0.325\textwidth]{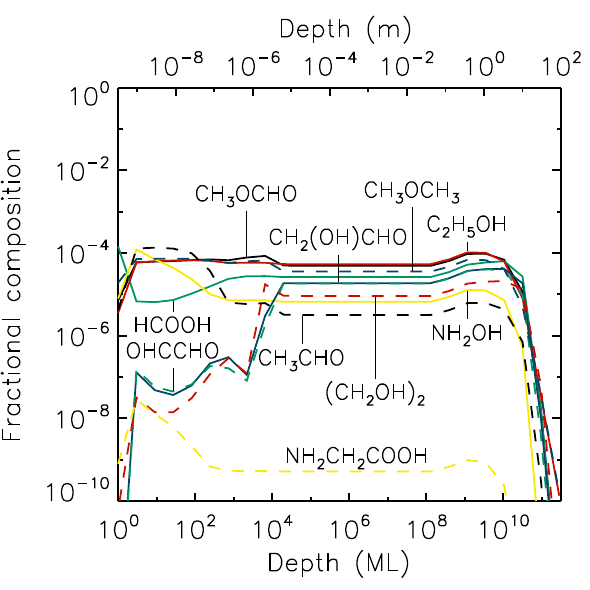}
    \caption{Fractional abundances of select ice species for the {\it second} of five solar approaches; Top: Abundances at second perihelion. Bottom: Abundances at second aphelion after starting point. Other information as per Fig.~\ref{hbSA1}}
    \label{hbSA2}
\end{figure*}

\begin{figure*}
    \centering
    \includegraphics[width=0.325\textwidth]{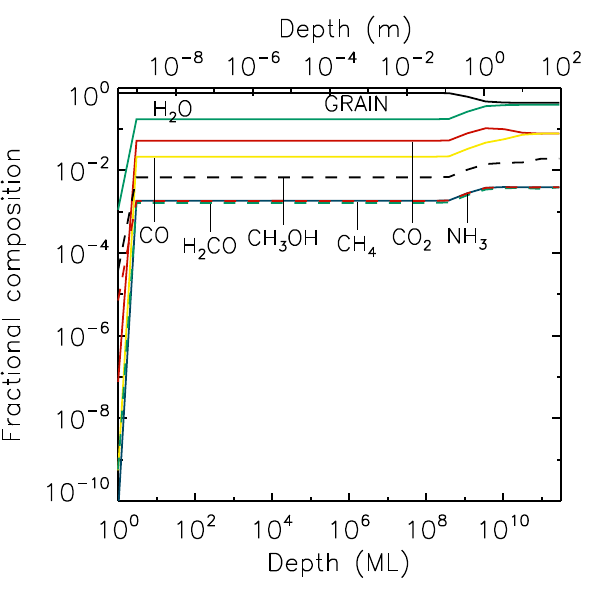}
    \includegraphics[width=0.325\textwidth]{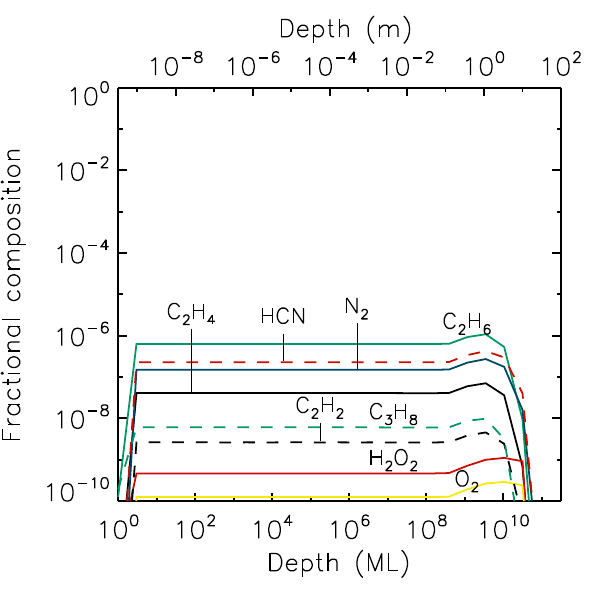}
    \includegraphics[width=0.325\textwidth]{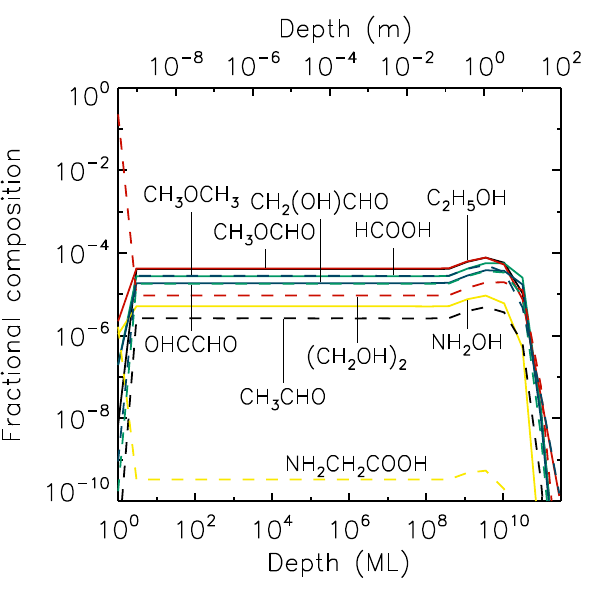}
    \\[\smallskipamount]
    \includegraphics[width=0.325\textwidth]{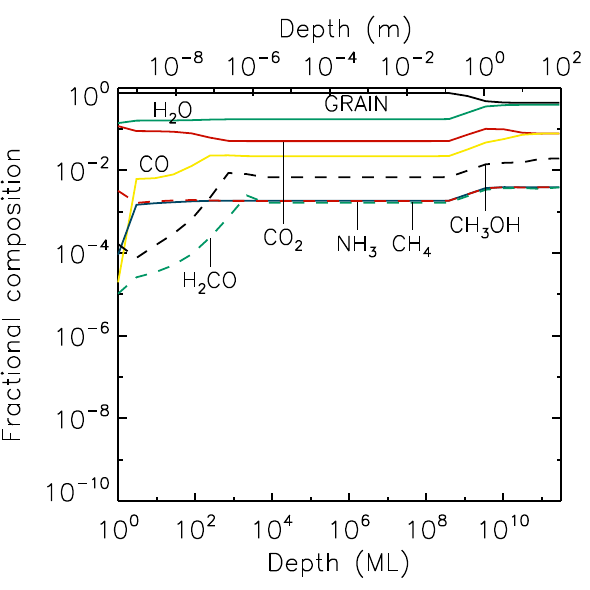}
    \includegraphics[width=0.325\textwidth]{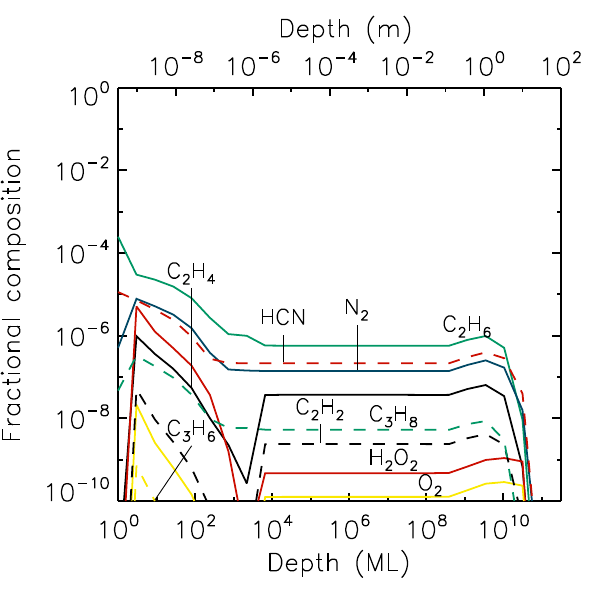}
    \includegraphics[width=0.325\textwidth]{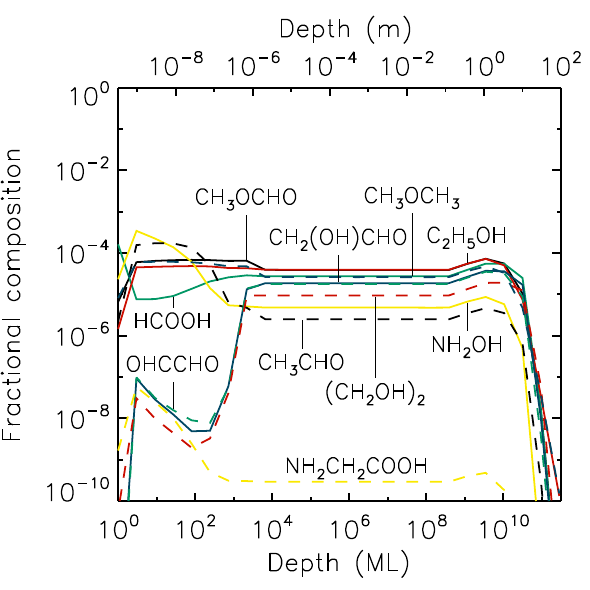}
    \caption{Fractional abundances of select ice species for the {\it final} of five solar approaches; Top: Abundances at fifth (final) perihelion. Bottom: Abundances at fifth (final) aphelion after starting point. Other information as per Fig.~\ref{hbSA1}}
    \label{hbSA5}
\end{figure*}

Here the simulated comet was allowed to follow the projected orbital path of comet Hale--Bopp; the chemical model results are shown in Figures \ref{hbSA1}-\ref{hbSA5}. As \ce{C2} and \ce{C3} are not nearly as abundant in the new model as in G19, and the solar approach phase does not appear to change this, two other species (\ce{C2H4} and \ce{C3H6}) are plotted instead. The species shown were similarly divided up into three different sets. Five sequential orbits for the comet were simulated (beginning at aphelion), based on the comet composition starting from that produced at the end of cold storage. Plots show the fractional abundances throughout the ice at perihelion and at the subsequent aphelion. "First aphelion" refers to the end of the first orbit rather than the starting point, and therefore "fifth/final aphelion" corresponds to the final time point. 

The first species set (left panels) are most useful for seeing the effects on the ice as a whole. At perihelion (upper panels), excluding the low abundances within the first few monolayers, the abundance profiles are very flat out to depths of \(10^{8}\) monolayers. Come aphelion (lower panels), some variation reminiscent of its cold storage phase appears again. With each subsequent orbit, the flat portion of the abundance profile edges further right, or deeper into the ice, decreasing in value for the ice and increasing for the grains. This general behavior, in which the dust becomes more concentrated in the upper layers while the surrounding ice becomes richer in the simple/primordial species stored deeper in the ice, is a direct result of the ice loss as a result of sublimation caused by close solar approach. The molecular content that was built up in those upper layers during the entirety of the cold storage phase has been lost, to be replaced (i.e at the same position relative to the new outer surface) by material originating deeper down.

Similar to the cold storage phase, by the time of the first aphelion (following the first solar approach), \ce{CO2} is enhanced while all other species in the set are diminished. This is refreshed each perihelion, whereby the profile flattens out before returning to a similar shape the following aphelion. While each individual cycle still seems to enhance \ce{CO2}, throughout the ice as a whole it is actually decreasing. In its place, the \ce{CO} abundance increases. 

Most other species tend overall to decrease over each subsequent orbit. The notable exceptions to this are \ce{O2} and \ce{H2O2} which increase slightly each perihelion. The overall fractional abundances of these two species increase over time, although their surface (upper layer) values at perihelion decrease instead. Some other species exhibit the opposite, with a net decrease in abundance deeper than 1 $\mu$m depth, yet a slight increase in the upper layers at aphelion. Examples of this include \ce{C2H6}, \ce{N2}, \ce{HCN}, \ce{NH2OH}, \ce{CH3CHO}, and \ce{HCOOH}. It is also worth noting that the first two orbits sometimes have species that do not follow a specific pattern, but in subsequent orbits seem to have adjusted and remain consistent. 

The third and fourth Solar Approach orbits are not shown in the figures, due to their similarity to the results to the fifth orbit, which is shown in Fig.~\ref{hbSA5}. In orbits 1 and 2 (Figs.~\ref{hbSA1} and \ref{hbSA2}), some species show irregular behavior, such as the low abundance of \ce{C2H2} and \ce{O2} in the upper micron during the first aphelion, which is then restored by the second aphelion, and the sudden dip of \ce{C2H4} at 1 $\mu$m during the second aphelion. 
By the third orbit, the cometary ice abundances remain consistent between comparable points in the orbits. While changes are still evident between perihelion and aphelion, as shown in Figure \ref{hbSA5}, the differences in abundances between perihelion three to five, as well as aphelion three to five, are negligible. This trend would presumably continue if further orbits were run.

Most COMs are formed in the cold storage phase through gradual hydrogenation and dehydrogenation of smaller species, followed by radical addition. During the active phase, COMs broadly maintain similar abundances throughout the ice. Following the first perihelion, most of the COMs in the outer layers of the comet are originally from deeper layers of the comet, having been formed originally during the cold storage phase by cosmic-ray radiolysis, and were left closer to the surface due to the mass-loss experienced during the comet's close solar passage. However, some COMs still form during the active phase, being driven by thermally activated reactions in response to the elevated temperatures. For example, methyl formate (CH$_3$OCHO) production is somewhat enhanced by the reaction CH$_3$O + CO $\rightarrow$ CH$_3$OCO, which is mediated by an activation-energy barrier of $\sim$33~kJ/mol (4000~K) as determined by gas-phase data \citep{Huynh2008}. The addition of atomic H to the CH$_3$OCO radical completes the formation of methyl formate.

Regarding smaller species, \ce{H2} and \ce{H} are the only species with mobility in the ice mantle within this model. By the second aphelion, the majority of the \ce{H2} budget has either reacted or moved to the surface and evaporated away. At all depths, \ce{O2} ice is primarily formed from dissociation of \ce{O2H}, as well as the direct addition of oxygen atoms following dissociation of species such as OH, which originate mostly from water. 

Once the inbound comet reaches about 5 AU it begins to experience significant loss of surface material due to thermal desorption, which outpaces all other formation and destruction routes. Similarly, once the outbound comet reaches 5 AU again, the surface loss has decreased below the other formation and destruction routes. Although there is indeed chemical activity within the ice during close solar approach, it is this mass loss, and the associated shift of deeper material toward the surface, that defines the major changes in chemical composition during the active-phase simulations. Post perihelion, the upper 1~$\mu$m of material again shows the effects of UV photo-processing. Given that much of this chemistry requires on the order of 1000 yr to reach maturity, as noted in \S~\ref{cs_results} above, the $\sim$2500 yr orbital period is sufficient for much of the same upper-layer composition to be re-formed by the return to aphelion.

\section{Discussion}
\label{discussion}

Before coming to the results of the chemical models, some discussion of the orbital parameter information for Hale--Bopp is warranted. As previously mentioned, the chemical model uses temperature profiles based on the heliocentric distance of comet Hale--Bopp over five orbits. Due to the uncertainty in the orbital behavior prior to its most recent solar approach, it was necessary to use the orbital parameters from the most recent orbit repetitively. Each of the five orbits in our model therefore uses this same set of positional data, corresponding to an orbital period of $\sim$2400 yr. However, there is evidence that Hale--Bopp's period has changed in the past; the previous apparition may have occurred $\sim$4200 yr before its most recent \citep{Marsden1997}. Assuming that the previous apparition was not its first, it is plausible that Hale--Bopp's prior orbital behavior could have been quite different from those two estimates as well.

While it is therefore unlikely, given the uncertainties, that our use of current orbital parameters is fully representative of Hale--Bopp's complete orbital history, the present, well-defined orbital parameters serve as a basis for obtaining from the model the broad chemical effects that Hale--Bopp is likely to have experienced over repeated active-phase orbits (assuming that it has indeed done so), or that it would do in future. It also serves as a basis for exploring the chemical behavior of the model itself under more stable conditions, before moving to models with more complex orbital dynamics in future work.

\S~\ref{model_results} makes note of chemical changes that occur at a depth of $\sim$1 $\mu$m into the ice. This can also be seen in many of the plots in Figures \ref{hb1e6}-\ref{hbSA5}. The interstellar UV field has a significant effect on the upper layers of the ice, but quickly tapers off at depths greater than about 1 $\mu$m. The significance of this particular depth is not arbitrary, but exists as a result of the limited UV penetration into the dust--ice mixture due to dust absorption. As a result, chemical changes are much more pronounced and diverse in these upper layers even from relatively early times in the models, allowing for the more varied and diverse complexity of molecules produced. The spikes and dips in the abundances mentioned throughout \S \ref{model_results} are a direct result of the decreasing influence of UV photons with depth.

Furthermore, many species also fall off in abundance at a depth of $\sim$10~m. This is a result of the limited penetration of Galactic cosmic rays, which induce their own chemistry to greater depths than UV, but with lower dissociation rates and thus a slower effect that requires the long period of the cold storage phase to produce significant amounts of product molecules. Beyond the penetration depth of GCRs, there remain pristine ice layers, which ultimately include the reservoir of material explained in detail in \S \ref{basic_meth}.  
A slight bump in many abundances occurs at the deepest abundances, peaking a little beyond the 1~m mark. This is caused by the slightly elevated dissociation rates caused by GCR activity (see G19, Fig.~11), which are in turn related to the shape of the GCR energy spectrum.

The models produce a range of interesting results. First of all, even though the models start off with no complex molecules, significant COM abundances ($\sim$\(10^{-5}\) with respect to water ice) can be formed during the cold-storage stage alone. These COMs are formed  primarily as a result of ($>$GeV) Galactic cosmic rays penetrating deep into the ice, which are present whether the comet is in the outer or the inner solar system. Galactic UV is also sufficient to enhance some complex ice species in the upper layers, while destroying others. 

These CR and UV effects persist throughout the solar approach phase as well, though now in tandem with a varying temperature profile driven by the solar heating. At the surface layers, many species notably drop in abundance during solar approach. This is not the case with some hydrocarbon abundances, which reach a peak in the upper layers. Some even grow in abundance with each subsequent orbit, indicating a gradual surface buildup of these hydrocarbons. 

Interestingly, molecular hydrogen abundances seem to fall off quickly after the comet enters the active phase. It takes only two active-phase orbits for \ce{H2} to have completely fallen off the scale at all layers of the ice (i.e. several orders of magnitude below a fraction $10^{-10}$ of total ice composition), and it never recovers. \ce{H2} and atomic hydrogen are the only bulk ice species allowed mobility within the ice. By this point, the entire budget of trapped H$_2$, formed largely by the production of atomic H by GCR dissociation of stable species and the subsequent recombination of those H atoms, has gone: used up either by its involvement in reactions or -- following accelerated diffusion through the ice driven by the higher temperatures -- desorption from the surface of the comet.

\begin{sidewaystable*}
    \centering
    \begin{tabular}{|c|c|c|c|c|c|c|c|c|c|c|c|c|}
        \hline
        Species & CS & Peri1 & Ap1 & Peri2 & Ap2 & Peri3 & Ap3 & Peri4 & Ap4 & Peri5 & Ap5 & Observational \\ \hline
        \ce{CO} & 1.7(-1) & 1.7(-1) & 1.7(-1) & 1.7(-1) & 1.7(-1) & 1.7(-1) & 1.7(-1) & 1.7(-1) & 1.7(-1) & 1.7(-1) & 1.8(-1) & 1.2-2.3(-1) \\ \hline
        \ce{CO2} & 2.4(-1) & 2.4(-1) & 2.4(-1) & 2.4(-1) & 2.4(-1) & 2.4(-1) & 2.3(-1) & 2.3(-1) & 2.3(-1) & 2.3(-1) & 2.3(-1) & 6(-2) \\ \hline
        \ce{CH4} & 1.0(-2) & 1.0(-2) & 1.0(-2) & 1.0(-2) & 1.0(-2) & 1.0(-2) & 1.0(-2) & 1.0(-2) & 1.0(-2) & 1.0(-2) & 1.0(-2) & 1.5(-2) \\ \hline
        \ce{H2CO} & 9.7(-3) & 9.7(-3) & 9.7(-3) & 9.7(-3) & 9.7(-3) & 9.7(-3) & 9.7(-3) & 9.7(-3) & 9.7(-3) & 9.7(-3) & 9.7(-3) & 1.1(-2) \\ \hline
        \ce{CH3OH} & 3.9(-2) & 3.9(-2) & 4.0(-2) & 4.0(-2) & 4.0(-2) & 4.0(-2) & 4.0(-2) & 4.0(-2) & 4.0(-2) & 4.0(-2) & 4.0(-2) & 2.4(-2) \\ \hline
        \ce{NH3} & 1.0(-2) & 1.0(-2) & 1.0(-2) & 1.0(-2) & 1.0(-2) & 1.0(-2) & 1.0(-2) & 1.0(-2) & 1.0(-2) & 1.0(-2) & 1.0(-2) & 7(-3) \\ \hline
        \ce{C2H2} & 4.7(-9) & 4.5(-9) & 4.2(-9) & 4.0(-9) & 3.7(-9) & 3.5(-9) & 3.3(-9) & 3.1(-9) & 2.9(-9) & 2.8(-9) & 2.6(-9) & 1-3(-3) \\ \hline
        \ce{C2H6} & 1.1(-6) & 1.1(-6) & 9.9(-7) & 9.4(-7) & 8.7(-7) & 8.2(-7) & 7.6(-7) & 7.2(-7) & 6.7(-7) & 6.3(-7) & 5.9(-7) & 6(-3) \\ \hline
        \ce{HCOOH} & 1.1(-4) & 1.1(-4) & 1.0(-4) & 1.0(-4) & 1.0(-4) & 1.0(-4) & 9.8(-5) & 9.7(-5) & 9.5(-5) & 9.3(-5) & 9.1(-5) & 9(-4) \\ \hline
        \ce{CH3CHO} & 6.4(-6) & 6.2(-6) & 6.0(-6) & 5.8(-6) & 5.5(-6) & 5.4(-6) & 5.2(-6) & 5.0(-6) & 4.8(-6) & 4.7(-6) & 4.5(-6) & 2(-4) \\ \hline
        \ce{CH3OCHO} & 1.0(-4) & 9.8(-5) & 9.4(-5) & 9.2(-5) & 8.8(-5) & 8.5(-5) & 8.2(-5) & 8.0(-5) & 7.6(-5) & 7.4(-5) & 7.1(-5) & 8(-4) \\ \hline
        \ce{CH3OCH3} & 6.2(-5) & 6.0(-5) & 5.7(-5) & 5.5(-5) & 5.2(-5) & 5.1(-5) & 4.8(-5) & 4.7(-5) & 4.4(-5) & 4.3(-5) & 4.1(-5) & $<$5(-3) \\ \hline
        \ce{(CH2OH)2} & 4.4(-5) & 4.3(-5) & 4.3(-5) & 4.2(-5) & 4.2(-5) & 4.1(-5) & 4.1(-5) & 4.0(-5) & 3.9(-5) & 3.9(-5) & 3.8(-5) & 2.5(-3) \\ \hline
        \ce{HCN} & 5.1(-7) & 5.0(-7) & 4.8(-7) & 4.6(-7) & 4.4(-7) & 4.3(-7) & 4.1(-7) & 3.9(-7) & 3.7(-7) & 3.6(-7) & 3.5(-7) & 2.5(-3) \\ \hline
        \ce{HNC} & 1.2(-8) & 1.2(-8) & 1.2(-8) & 1.2(-8) & 1.2(-8) & 1.1(-8) & 1.1(-8) & 1.1(-8) & 1.1(-8) & 1.1(-8) & 1.1(-8) & 4(-4) \\ \hline
        \ce{HNCO} & 3.4(-6) & 3.4(-6) & 3.4(-6) & 3.4(-6) & 3.4(-6) & 3.3(-6) & 3.3(-6) & 3.3(-6) & 3.3(-6) & 3.3(-6) & 3.2(-6) & 1(-3) \\ \hline
        \ce{CH3CN} & 6.6(-15) & 6.2(-15) & 5.7(-15) & 5.3(-15) & 4.7(-15) & 4.4(-15) & 3.9(-15) & 3.6(-15) & 3.3(-15) & 3.0(-15) & 2.7(-15) & 2(-4) \\ \hline
        \ce{HC3N} & 5.8(-19) & 5.6(-19) & 5.3(-19) & 5.1(-19) & 4.8(-19) & 4.7(-19) & 4.4(-19) & 4.2(-19) & 4.0(-19) & 3.8(-19) & 3.6(-19) & 2(-4) \\ \hline
        \ce{NH2CHO} & 2.7(-7) & 2.7(-7) & 2.7(-7) & 2.7(-7) & 2.6(-7) & 2.6(-7) & 2.6(-7) & 2.6(-7) & 2.6(-7) & 2.5(-7) & 2.5(-7) & 1.5(-4) \\ \hline
    \end{tabular}
    \caption{Integrated fractional abundances (with respect to water) of selected chemical species, to a depth of 15~m, for various times in the models. Gas-phase observational values, shown for comparison, are taken from \citet{Mumma2011}. Values are indicated in the form $A(B) = A \times 10^B$. `CS' denotes the end of the cold storage phase (4.5~Gyr); `Peri' stands for perihelion, and `Ap' for aphelion, with the orbit number also indicated.}
    \label{hb_abuns}
\end{sidewaystable*}

One of the most notable molecules in the chemical network is the amino acid glycine (\ce{NH2CH2COOH}), due to its significance to life, as well as recent cometary detections. Despite the ability of \emph{MAGICKAL} to generate significant COM abundances within the simulated cometary nucleus, glycine still remains at relatively low abundances (versus the $\sim$1\% with respect to water observed in comet 67P). Somewhat higher abundance can be seen in the upper micron, which reaches a peak during the cold storage phase. However, much of this upper-layer abundance is lost during solar approach. 

As noted earlier, in \S \ref{sa_results}, comparison of the same points in the first two orbits does not demonstrate a clear trend in the evolution of all chemical species at all depths, and some species seem somewhat unstable in those earlier orbits, particularly when comparing between aphelions. It is not until the third orbit that the ice abundances seem to stabilize and follow a slow and steady trend over time. 
Each individual orbit is identical in terms of orbital parameters, and while the temperature profiles have small variations between orbits that are essentially gone after the second orbit, these variations are small enough not to be a likely cause for substantial chemical variation between orbits.

What may be seen, however, between the end-time abundance profiles of the cold-storage model and the perihelion and aphelion profiles for the first orbital model, is the extension of the deep-ice abundances (of COMs and other species) toward the upper layers (i.e. shifting from right to left); the desorption of material from the surface during the extreme heating of the comet around the time of perihelion leads to the loss of much of the material in at least the upper 1~cm, to be replaced at those depths with material that was heavily processed by GCRs during cold storage, but which was not subject to UV processing. Following the loss of what were previously the upper layers, the remaining material is now exposed to interstellar UV, so that by aphelion it has undergone some of the same processing as the original material did.

When the model comet embarks on its second passage toward the sun, the new set of upper layers that were revealed by solar processing in the previous orbit, and that were chemically processed by interstellar UV, are then subject to the same thermally induced loss. They are then replaced by new material that is subsequently processed in the same way. This new material will be much more similar in its initial composition to the material that it directly replaced, as compared with the upper-layer material that was directly inherited from the cold storage phase. Subsequent solar orbits would bring new material to the surface that shows the same behavior. 

It can therefore be seen that, in these models, the material that is distinctly associated with interstellar UV during cold storage is lost to space during the first solar approach. However, the material that will be lost during the second solar approach will be material produced by GCRs at much greater depths (up to around 10~m or so). Furthermore, the subsequent passages do not remove enough material to deplete this GCR-processed material. Based on the total loss of ice material in units of monolayers, the comet loses $\sim$19.1~cm of ice from its surface per solar orbit, which remains consistent within a few percent for each orbit. For comparison, the total ice lost during the full cold storage phase is $\sim$19.3~cm. Although this means that the upper micron of material is lost many times over during even the cold phase, the loss rate of 1~$\mu$m per $\sim$23,000 yr means that the chemistry down to 1~$\mu$m is able to respond relatively rapidly, due to its characteristic timescale on the order of 1000~yr. The chemistry in those upper layers therefore remains quite stable during cold storage, in spite of the losses. The only substantial change in composition on long timescales is the gradual enrichment of the dust component.

It should be noted that the mass-loss discussed above corresponds specifically and uniquely to sublimation of the ice from the surface, and not from outbursts, which involves 
structural changes from cliff collapse and bursting of voids under pressure. The model is not presently capable of incorporating such a rapid and structurally complex occurrence. The ice that is modeled here should therefore be considered representative of regions of the comet that are left unaffected by any putative local outburst events.

\subsection{Comparison with previous modeling techniques}
\label{model_comp}

In a broad sense, the new chemical results for cold storage are similar to those presented by G19; however, it should be noted that the G19 model with which we have compared our new data corresponds specifically to the 10~K case. As discussed in depth in that paper, the elevation of the temperature in the old model to 20~K (in all layers) induces bulk diffusion of large species that is not seen in the new models at any temperature, due to explicit changes in modeling techniques. Nevertheless, for comparable 10~K models, more appropriate to an Oort Cloud object, the differences in the chemistry between the new and old treatments are a matter of degree rather than a change in character.

Although not strongly manifested in the cold storage simulations, some of the updates to the chemical model are highly necessary in order to simulate the later solar approach. This is particularly true for the new treatment of back-diffusion within the ice when considering the rates of transfer between discrete layers. Although bulk diffusion in the new model is now strictly limited to H and H$_2$, these two species become highly diffusive in the bulk ice as temperatures increase. Meanwhile, as that diffusion occurs, these species will make their way to the surface of the comet more rapidly, allowing them to escape. The findings of the Monte Carlo simulations of three-dimensional random walk, and their parameterization into manageable expressions, indicate that the rates of transfer should be strongly affected by the back-diffusion effect -- far more than any comparable effect related to reactions, either in the bulk ice or on the ice surface. The effect on these models is to allow the modest amounts of H and H$_2$ that exist in the deepest layers to be retained for longer, allowing, in particular, a somewhat more efficient process of H-abstraction and hydrogenation of radicals to occur deep in the ice. The adoption by G19 of simple diffusion rates between layers, corresponding to the high-occupation case, is clearly inappropriate for the more typical low-occupation case for H and H$_2$. For example, in the deepest layer in the model, which has a thickness of $\sim$$2.8 \times 10^{11}$ ML ($\sim$90~m), the low-occupation case provides a back-diffusion factor of around $10^{23}$; an enormous modification to the rate of loss of H and H$_2$ by diffusion between layers compared with the simple, high-occupation rate.

It is indeed also notable that back diffusion has only a very limited effect on rates of reaction within the bulk ice; the effect is essentially negligible, being typically less than a factor 1.5, which is smaller than any uncertainty on the fundamental diffusion rates. Thus, while H and H$_2$ are much more effectively retained in the ice due to their large random walk, their efficiency in finding reaction partners in three dimensions is effectively unimpeded.

Our calculation of both these quantities for conditions appropriate to interstellar dust-grain ice mantles is also valuable, and allows the degree of occupation to be taken into account in the bulk-ice diffusion rates for those much thinner ices.

\begin{sidewaystable*}
    \centering
    \begin{tabular}{|c|c|c|c|c|c|c|c|c|c|c|c|c|}
        \hline
        Species & CS & Peri1 & Ap1 & Peri2 & Ap2 & Peri3 & Ap3 & Peri4 & Ap4 & Peri5 & Ap5 & Observational \\ \hline
	\ce{CO}	&	 1.0(-1)	 & 	 1.0(-1)	 &	 1.1(-1)	 &	 1.1(-1)	 &	 1.1(-1)	 &	 1.2(-1)	 &	 1.2(-1)	 &	 1.2(-1)	 &	 1.3(-1)	 &	 1.3(-1)	 &	 1.3(-1)	&	1.2-2.3(-1)	\\	\hline
	\ce{CO2}	&	 3.3(-1)	 & 	 3.3(-1)	 &	 3.3(-1)	 &	 3.2(-1)	 &	 3.2(-1)	 &	 3.1(-1)	 &	 3.1(-1)	 &	 3.0(-1)	 &	 3.0(-1)	 &	 3.0(-1)	 &	 2.9(-1)	&	6(-2)	\\	\hline
	\ce{CH4}	&	 1.1(-2)	 & 	 1.1(-2)	 &	 1.1(-2)	 &	 1.1(-2)	 &	 1.1(-2)	 &	 1.1(-2)	 &	 1.1(-2)	 &	 1.1(-2)	 &	 1.1(-2)	 &	 1.1(-2)	 &	 1.1(-2)	&	1.5(-2)	\\	\hline
	\ce{H2CO}	&	 9.2(-3)	 & 	 9.2(-3)	 &	 9.2(-3)	 &	 9.3(-3)	 &	 9.3(-3)	 &	 9.4(-3)	 &	 9.5(-3)	 &	 9.5(-3)	 &	 9.5(-3)	 &	 9.6(-3)	 &	 9.6(-3)	&	1.1(-2)	\\	\hline
	\ce{CH3OH}	&	 3.8(-2)	 & 	 3.9(-2)	 &	 3.9(-2)	 &	 3.9(-2)	 &	 3.9(-2)	 &	 3.9(-2)	 &	 3.9(-2)	 &	 3.9(-2)	 &	 4.0(-2)	 &	 4.0(-2)	 &	 4.0(-2)	&	2.4(-2)	\\	\hline
	\ce{NH3}	&	 1.1(-2)	 & 	 1.1(-2)	 &	 1.1(-2)	 &	 1.1(-2)	 &	 1.1(-2)	 &	 1.1(-2)	 &	 1.1(-2)	 &	 1.1(-2)	 &	 1.1(-2)	 &	 1.1(-2)	 &	 1.1(-2)	&	7(-3)	\\	\hline
	\ce{C2H2}	&	 2.4(-8)	 & 	 2.3(-8)	 &	 2.2(-8)	 &	 2.1(-8)	 &	 1.9(-8)	 &	 1.8(-8)	 &	 1.7(-8)	 &	 1.6(-8)	 &	 1.5(-8)	 &	 1.4(-8)	 &	 1.3(-8)	&	1-3(-3)	\\	\hline
	\ce{C2H6}	&	 5.9(-6)	 & 	 5.8(-6)	 &	 5.4(-6)	 &	 5.2(-6)	 &	 4.7(-6)	 &	 4.5(-6)	 &	 4.1(-6)	 &	 3.8(-6)	 &	 3.5(-6)	 &	 3.3(-6)	 &	 3.0(-6)	&	6(-3)	\\	\hline
	\ce{HCOOH}	&	 1.5(-4)	 & 	 1.5(-4)	 &	 1.5(-4)	 &	 1.5(-4)	 &	 1.6(-4)	 &	 1.6(-4)	 &	 1.6(-4)	 &	 1.6(-4)	 &	 1.6(-4)	 &	 1.6(-4)	 &	 1.6(-4)	&	9(-4)	\\	\hline
	\ce{CH3CHO}	&	 2.0(-5)	 & 	 1.9(-5)	 &	 1.9(-5)	 &	 1.8(-5)	 &	 1.7(-5)	 &	 1.7(-5)	 &	 1.6(-5)	 &	 1.6(-5)	 &	 1.5(-5)	 &	 1.4(-5)	 &	 1.4(-5)	&	2(-4)	\\	\hline
	\ce{CH3OCHO}	&	 3.0(-4)	 & 	 3.0(-4)	 &	 2.9(-4)	 &	 2.8(-4)	 &	 2.7(-4)	 &	 2.6(-4)	 &	 2.5(-4)	 &	 2.4(-4)	 &	 2.3(-4)	 &	 2.2(-4)	 &	 2.1(-4)	&	8(-4)	\\	\hline
	\ce{CH3OCH3}	&	 2.2(-4)	 & 	 2.2(-4)	 &	 2.1(-4)	 &	 2.0(-4)	 &	 1.9(-4)	 &	 1.9(-4)	 &	 1.7(-4)	 &	 1.7(-4)	 &	 1.6(-4)	 &	 1.5(-4)	 &	 1.4(-4)	&	$<$5(-3)	\\	\hline
	\ce{(CH2OH)2}	&	 5.3(-5)	 & 	 5.3(-5)	 &	 5.4(-5)	 &	 5.4(-5)	 &	 5.4(-5)	 &	 5.4(-5)	 &	 5.5(-5)	 &	 5.5(-5)	 &	 5.4(-5)	 &	 5.4(-5)	 &	 5.4(-5)	&	2.5(-3)	\\	\hline
	\ce{HCN}	&	 1.8(-6)	 & 	 1.7(-6)	 &	 1.7(-6)	 &	 1.6(-6)	 &	 1.6(-6)	 &	 1.5(-6)	 &	 1.4(-6)	 &	 1.4(-6)	 &	 1.3(-6)	 &	 1.2(-6)	 &	 1.2(-6)	&	2.5(-3)	\\	\hline
	\ce{HNC}	&	 7.2(-9)	 & 	 7.3(-9)	 &	 7.5(-9)	 &	 7.7(-9)	 &	 8.1(-9)	 &	 8.3(-9)	 &	 8.6(-9)	 &	 8.8(-9)	 &	 9.1(-9)	 &	 9.3(-9)	 &	 9.5(-9)	&	4(-4)	\\	\hline
	\ce{HNCO}	&	 2.1(-6)	 & 	 2.1(-6)	 &	 2.2(-6)	 &	 2.2(-6)	 &	 2.3(-6)	 &	 2.4(-6)	 &	 2.5(-6)	 &	 2.5(-6)	 &	 2.6(-6)	 &	 2.7(-6)	 &	 2.7(-6)	&	1(-3)	\\	\hline
	\ce{CH3CN}	&	4.6(-14)	 & 	4.5(-14)	 &	4.2(-14)	 &	3.9(-14)	 &	3.5(-14)	 &	3.2(-14)	 &	2.9(-14)	 &	2.6(-14)	 &	2.3(-14)	 &	2.1(-14)	 &	1.9(-14)	&	2(-4)	\\	\hline
	\ce{HC3N}	&	2.4(-18)	 & 	2.3(-18)	 &	2.3(-18)	 &	2.2(-18)	 &	2.1(-18)	 &	2.0(-18)	 &	1.9(-18)	 &	1.8(-18)	 &	1.7(-18)	 &	1.6(-18)	 &	1.5(-18)	&	2(-4)	\\	\hline
	\ce{NH2CHO}	&	 2.2(-7)	 & 	 2.1(-7)	 &	 2.1(-7)	 &	 2.2(-7)	 &	 2.2(-7)	 &	 2.2(-7)	 &	 2.3(-7)	 &	 2.3(-7)	 &	 2.4(-7)	 &	 2.4(-7)	 &	 2.4(-7)	&	1.5(-4)	\\	\hline
    \end{tabular}
    \caption{Integrated fractional abundances (with respect to water) of selected chemical species, to a depth of 1~m, for various times in the models. Gas-phase observational values, shown for comparison, are taken from \citet{Mumma2011}. Values are indicated in the form $A(B) = A \times 10^B$. `CS' denotes the end of the cold storage phase (4.5~Gyr); `Peri' stands for perihelion, and `Ap' for aphelion, with the orbit number also indicated.}
    \label{hb_abuns_2}
\end{sidewaystable*}

\subsection{Comparison with Hale--Bopp abundances}
\label{hb_comp}

Although we have adopted orbital parameters for the solar approach phase corresponding directly to Hale--Bopp, the chemical model outputs have not so far been directly compared with those of the comet itself. As noted in \S~\ref{discussion}, the present model cannot and does not attempt to simulate outburst events. However, following the approach of G19, it is useful to compare the solid-phase abundances produced by the model at various points in its dynamical/physical evolution with the gas-phase abundances determined for Hale--Bopp. To allow such a comparison, the fractional abundances of selected ice species are integrated to a depth of $\sim$15~m into the model comet, corresponding to the maximum reach of GCR processing. These abundances are presented as a fraction of integrated water ice in Table~\ref{hb_abuns}, with values provided at the end of the cold storage phase and at each of the perihelion and aphelion points of the five simulated active-phase orbits. Observational abundances collated by \citet{Mumma2011} are also shown. 

Three of the listed species, \ce{CH4}, \ce{H2CO}, and \ce{NH3}, which are those included in the initial composition of the comet at the beginning of the cold storage phase, remain consistent throughout the model runs. Two other species, \ce{CO} and \ce{CH3OH}, increase slightly. All other species exhibit gradual decreases, yet at differing rates. It should be noted that all the species that either increase or remain constant are species in the initial set, and with the exception of \ce{CO2} none of the initial species decrease. In each case, the fractional abundances are somewhat linear and experience no oscillatory behavior associated with varying orbital distance and the related temperature changes. Meanwhile, the effect of solar approach on COM abundances is a universal net negative. In the case of molecules not present in the initial, pristine ice composition, each solar approach reduces the overall inventory of all ice species. 

When compared to the observational values, all of the starting species with the exception of \ce{CO2} are within a factor of 2 or less. The \ce{CO2} abundance is around 4x above the observed value. Most of the \ce{CO2} in the ice derives from the initial abundance set at the beginning of the cold storage phase, although there is some conversion from CO and \ce{H2O} occurring mainly during cold storage. The discrepancy may therefore be due to the choice of initial abundance.
Almost all other species shown in Table~\ref{hb_abuns} appear to be several orders of magnitude below observational values. \ce{HCOOH} and \ce{CH3OCHO} are the only ice species roughly within an order of magnitude of the observational values, while \ce{CH3CHO} and \ce{(CH2OH)2} are roughly within two orders of magnitude. It is also worth mentioning that for \ce{CH3OCH3} there is only an observational upper limit, and our computational value lies roughly two orders of magnitude below this value. 

It is interesting to note that the modeled ratios of the oxygen-bearing COMs \ce{HCOOH}, \ce{CH3OCHO} and \ce{CH3CHO} relative to each other are reasonably consistent with the observed values, with the abundance of \ce{CH3OCH3} also being technically consistent with the observed upper limit. In contrast, the nitrogen-bearing molecules in the Table, aside from \ce{NH3} (which is one of the initial ice species), are all underestimated by at least three orders of magnitude. Similarly, the organic molecules which do not contain oxygen similarly underestimate observational values by at least four orders of magnitude. 

Overall, the disparity between the model and observational values is not significantly impacted by the processing from the solar approach phase. Despite the fact that all of the non-starting species decrease with time, the orders of magnitude differences between the model and observational values make this relatively insignificant. Even values from the end of the cold storage phase, when the fractional abundances of non-starting species are the highest, still notably underestimate all observational values. 

The integration of local abundances to 15~m depth follows the approach of G19; however, it is noteworthy that in the present (cold storage) model, the abundances of some species begin to decline at depths less than this maximum value. In the G19 models, a small degree of thermal diffusion (even at 10~K) especially for small species may have allowed abundances to be smoothed out somewhat at around the integration threshold. Here, however, no such mechanism exists. Furthermore, due to the exponentially increasing depth of each distinct chemical layer in the model, the abundances in deeper layers naturally contribute more to the integrated values than those of the upper layers. To ensure that the comparison with observational data is fair, we present in Table~\ref{hb_abuns_2} the integrated values to a depth of 1~m total. This value also corresponds approximately to the total amount of material lost from the surface over the five orbits.

Using 1~m integration depth, most simple species nevertheless remain at fairly comparable values with the 15~m integration, although the changes are still significant, as the strongest depth-dependent variation for these species occurs at or around the 1~m mark. This is mainly because the loss of material extends to a depth of this order, becoming greater with each orbit (by around 19~cm). This means that the region of concentrated dust also extends to these depths.

For species not in the initial set, such as the hydrocarbons larger than methane, there is a modest increase in the ratio with respect to water. Much of this is caused by the drop in the integrated water content, by a factor 2.25 for the final aphelion. A number of species' abundances increase by a factor greater than this, indicating that they did not extend fully to 15~m. However, the changes are at most a further factor of around 2, and some species record a slight fall for the 1~m integration. In spite of this, a few of the COMs ultimately come a little closer to the observed values, e.g. methyl formate (\ce{CH3OCHO}); thus, according to the model, at least a substantial fraction of the methyl formate observed in the gas phase could plausibly have an origin in GCR-induced processing during the cold storage. But aside from dimethyl ether, the rest are rather too low.

Table~\ref{hb_now_vs_g19} directly compares the updated model abundances versus those in G19. The table displays the abundances at the end-time of 4.5~Gyr for both models, using an integration depth of 15m and 1m. Additionally, we identify the fractional change from G19 to the current model. Many ice species remain relatively consistent between the models, especially those included in our initial composition. Some undertake rather drastic decreases such as \ce{C2H2}, \ce{CH3CN}, and \ce{HC3N} while some of the others have smaller, yet still significant, decreases. This general decrease in abundances can again be largely explained by the absence of bulk diffusion for species aside from atomic and molecular hydrogen as noted above.

Finally, it is instructive to consider how much of each molecule in the comparison has been ejected from the surface (intact) during each stage of the chemical evolution. This may be determined simply by summing the amounts of each molecule desorbed in various ways between each key time-point (e.g. perihelion to aphelion), as shown in Table~\ref{hb_delta_gas}. Here, we see that the material released into the gas phase during the solar approach phases is fairly consistent with the composition of the processed ice, more particularly with the upper 1~m integrated values. In general, the solar heating and desorption therefore appears to release the various ice components in a fairly uniform fashion. The surface temperatures achieved at peak should indeed be sufficient to allow the various species to desorb freely once they reach the upper surface layer. The exception is ethylene glycol, whose abundance in the gas with respect to water in each solar approach stage is a factor of $\sim$5 higher, and falls within an order of magnitude of the observed value. We note that our comparisons of course do not consider any gas-phase chemistry that could enhance or diminish molecular abundances following release either through the desorption mechanisms considered here or by localized outburst events.

It remains, however, a substantial challenge to construct a framework in which the outburst process may be meaningfully incorporated into this already quite complex chemical treatment of comet nucleus chemistry. Similarly, it is by no means trivial to consider the process of diffusion on internal pore surfaces within the ice (to allow transport between different depths), nor to include the associated surface reactions and the growth (or collapse) of the pores over time.

Another aspect that is also not considered in the models is the possible loss of dust material as the result of gas drag from the continuous sublimation of volatiles over the entire surface. Modeling this process accurately would require a deeper analysis of the mechanics of dust loss than can be considered in the present work. However, dust loss through this mechanism could make a significant contribution to the overall mass loss from a comet, thus affecting our comparisons with observationally derived mass-loss estimates. There is thus room for further development of the model in this direction.

A more immediately tractable process for inclusion in the models that has not been considered here is the influence on the chemistry of UV photons and energetic protons originating from the Sun itself, which may be able to compete with interstellar photons in their processing of the upper ice layers for small heliocentric distances. The planned follow-up to the present publication (i.e. paper II) will address these additional drivers of chemical evolution. Also, the use of a more explicit calculation of initial ice abundances, derived from interstellar chemical models, may provide a more accurate starting point for the cold-storage ice composition, by including atoms, radicals, and sparse molecular components that could affect the later bulk-ice chemical behavior and physical structure. This would also provide a way to test more directly the survival of complex organic molecules inherited from earlier stages of solar-system evolution.

Based on these four methods of comparison of modeled ice abundances with the available observational data, the quantities of product species, including COMs, produced during the cold storage and active phases are insufficient to account for the majority of observed abundances in the gas phase, excluding the simple, abundant ice constituents that are present in each layer from the beginning of the model. It is nevertheless interesting that certain quantities of COMs and other species, while not great enough to meet observational targets, are likely to be enhanced by the cold storage stage in particular, due to GCR processing. 

The good match for the simpler species, such as CO, CH$_3$OH, etc., is a direct result of our choice of initial ice composition. The initial ice composition is already similar to interstellar ice compositions by design. All other species ultimately fail to be reproduced through methods currently implemented into the model. As noted by G19, for new comets the chemical signatures produced by the release of nucleus material could be somewhat different, and chemically richer, than those of comets that have undergone multiple passes and have thus already lost their outer $\sim$10~m of ice.

As to the origins of the observed COMs, given their abundances, these models would indicate that those molecules must have been incorporated into the comets from pre-existing material. This is a rather stronger conclusion than that provided by G19, who found some species could plausibly reach observable abundances to $\sim$15~m depths. However, as the present models do not yet include anything other than a limited selection of simple, initial species, it is unclear whether having a broader selection of initial species could still lead to some significant in situ production of larger species. It is also unknown to what degree any COMs inherited from primordial material might be diminished, rather than enhanced, by cosmic ray and/or UV processing.

\subsection{Mass loss from Hale--Bopp}
\label{hb_loss}

While not implemented as a function of the model, it is worth considering the effects of local outbursts of cometary nucleus material. Adding such effects into our models would cause further loss of ice species in the upper layers -- including the dust itself, which is not allowed to leave the surface of the comet in the present model. As shown by Figs.~\ref{hb4e9}--\ref{hbSA5}, the abundances of complex ice species fall off drastically at depths greater than around 10--15~m. If, during the active phase, the comet were to lose its top 15~m of ice due to outbursts, the inventory of COMs formed during cold storage would be substantially depleted, replaced with ice material from the unprocessed core, and resulting in an almost completely pristine ice. From there, the remaining inventory could be entirely depleted following a second orbit, leaving behind a pristine ice.

It is worthwhile to consider the observational estimates of mass-loss from Hale--Bopp. \citet{Jewitt1999} measured the sub-mm dust continuum emission in the coma, determining the total dust-mass lost during the 1997 apparition to be $\sim$$3 \times 10^{13}$~kg. Thence, they determined the ratio of dust- to water-mass production rates to be $>$5. Those authors then assumed that the water and dust loss could be attributed to active vents on the surface. Assuming a lower limit for the exposed area responsible for water production of 2800~km$^2$, and combining this with an assumed bulk density of 1~g~cm$^{-3}$, they determined that the vent floors should have receded by a depth of $\sim$10~m per orbit.

As previously noted, our model displays a mass loss for Hale--Bopp of around 19~cm per orbit; with our adopted water-ice density of 0.9~g~cm$^{-3}$, this corresponds to an ice-loss per orbit of 172~kg~m$^{-2}$. The water-only mass loss is 112~kg~m$^{-2}$. If we allow, in our one-dimensional treatment, that all regions of the surface are actively able to sublimate, then a surface area of $\sim$11,300~km$^{2}$ (assuming $r=30$~km) provides a total water-mass loss per orbit of $1.3 \times 10^{12}$ kg and a total ice-mass loss per orbit of $1.9 \times 10^{12}$ kg. On this basis, the model is reasonably consistent with the observationally determined total water loss of $<6 \times 10^{12}$~kg.

Although our model does not currently implement dust loss mechanisms, we may further estimate how much dust might be lost along with the sublimating water. After the upper layers have been depleted of much of their ice content, which occurs even during the cold storage phase, our model produces a dust- to water-ice ratio of 5.37 by volume within the upper layers. Assuming that this ratio is maintained as dust and ice are lost from the surface, and assuming a dust density of 3~g~cm$^{-3}$, our water-loss rate would indicate a total dust-loss per orbit of $2.3 \times 10^{13}$ kg. Therefore, on the basis of a fully active surface for Hale--Bopp, both our explicit water-loss rate and the implied dust-loss rate align well with the values determined by \citet{Jewitt1999}.

However, since this dust-loss process is not included in the model, a larger {\em depth} of material would be lost than our current model presumes. Since the maximum fractional dust content per layer allowed in our model (based on geometry) is $\sim$74\%, approximately 73~cm of total material would be lost from the entire (non-localized) surface per orbit. While this is still a substantial difference, it is not sufficient to remove even the majority of the 10--15~m of processed ice over the five orbits tested here in our model. A less efficient removal of grains from the surface would also produce a smaller discrepancy in the depth of material lost between these two modeling cases, although this would inevitably lead to a poorer match to the dust-production rate.

Given that the approximate water-loss rate of the model corresponds to the observational value, we can examine two extreme scenarios regarding the active area of surface mass-loss. \citet{Jewitt1999} assume a very localized mass loss, corresponding to an active region of only a quarter of the available surface, based on a comet radius of 30~km. Under their scenario, the vent regions would go deep, exposing pockets of pristine ice. Meanwhile, the remaining regions of the surface with low activity would presumably still preserve the material that our models suggest to be enriched by UV and GCR processing during cold storage. In such a case, while our models would appear to underestimate the amount of material lost in those vent regions, they would overestimate the loss from the inactive region.

Alternatively, adopting the fully active surface implicitly considered by our model, while the upper micron of ice initially processed by the interstellar UV field is lost, a substantial amount of the ice processed by GCRs is retained even after five orbits. The reality, however, is likely some middle ground between these two scenarios. For example, our model cannot presently distinguish between surface regions that may experience different amounts of solar heating during the active phase. The lack of an explicit dust-loss mechanism is also an area where the model could be improved.

\begin{table*}
    \centering
    \begin{tabular}{|c|c|c|c|c|c|c|}
        \hline
        Species & New (15m) & G19 (15m) & New/G19 (15m) & New (1m) & G19 (1m) & New/G19 (1m) \\ \hline
        \ce{CO} & 1.7(-1) & 1.5(-1) & 1.1(+0) & 1.0(-1) & 7.8(-2) & 1.3(+0) \\ \hline
        \ce{CO2} & 2.4(-1) & 2.3(-1) & 1.0(+0) & 3.3(-1) & 2.9(-1) & 1.1(+0) \\ \hline
        \ce{CH4} & 1.0(-2) & 1.8(-2) & 5.6(-1) & 1.1(-2) & 3.5(-2) & 3.1(-1) \\ \hline
        \ce{H2CO} & 9.7(-3) & 1.2(-2) & 8.1(-1) & 9.2(-3) & 1.3(-2) & 7.1(-1) \\ \hline
        \ce{CH3OH} & 3.9(-2) & 2.2(-2) & 1.8(+0) & 3.8(-2) & 1.1(-2) & 3.5(+0) \\ \hline
        \ce{NH3} & 1.0(-2) & 9.0(-3) & 1.1(+0) & 1.1(-2) & 6.1(-3) & 1.8(+0) \\ \hline
        \ce{C2H2} & 4.7(-9) & 9.1(-5) & 5.2(-5) & 2.4(-8) & 4.7(-4) & 5.1(-5) \\ \hline
        \ce{C2H6} & 1.1(-6) & 3.8(-4) & 2.9(-3) & 5.9(-6) & 2.0(-3) & 3.0(-3) \\ \hline
        \ce{HCOOH} & 1.1(-4) & 2.5(-4) & 4.4(-1) & 1.5(-4) & 5.9(-4) & 2.5(-1) \\ \hline
        \ce{CH3CHO} & 6.4(-6) & 2.5(-4) & 2.6(-2) & 2.0(-5) & 7.1(-4) & 2.8(-2) \\ \hline
        \ce{CH3OCHO} & 1.0(-4) & 2.8(-4) & 3.6(-1) & 3.0(-4) & 9.1(-4) & 3.3(-1) \\ \hline
        \ce{CH3OCH3} & 6.2(-5) & 5.4(-5) & 1.2(+0) & 2.2(-4) & 2.1(-4) & 1.1(+0) \\ \hline
        \ce{(CH2OH)2} & 4.4(-5) & 4.6(-5) & 9.6(-1) & 5.3(-5) & 7.7(-5) & 6.9(-1) \\ \hline
        \ce{HCN} & 5.1(-7) & 9.9(-5) & 5.2(-3) & 1.8(-6) & 5.5(-4) & 3.3(-3) \\ \hline
        \ce{HNC} & 1.2(-8) & 7.4(-5) & 1.6(-4) & 7.2(-9) & 1.6(-4) & 4.5(-5) \\ \hline
        \ce{HNCO} & 3.4(-6) & 1.2(-5) & 2.8(-1) & 2.1(-6) & 2.0(-5) & 1.1(-1) \\ \hline
        \ce{CH3CN} & 6.6(-15) & 1.2(-5) & 1.7(-8) & 4.6(-14) & 2.3(-6) & 2.0(-8) \\ \hline
        \ce{HC3N} & 5.8(-19) & 5.2(-9) & 1.1(-10) & 2.4(-18) & 2.2(-8) & 1.1(-10) \\ \hline
        \ce{NH2CHO} & 2.7(-7) & 2.2(-5) & 1.2(-2) & 2.2(-7) & 4.4(-5) & 5.0(-3) \\ \hline
    \end{tabular}
    \caption{Comparison of the new results and the G19 model outputs at the endpoint of the cold-storage stage, based on the integrated fractional abundances of selected chemical species to a depth of 15~m and 1~m. The ratios of the new results versus G19 are also listed for each integration depth. Values are indicated in the form $A(B) = A \times 10^B$.}
    \label{hb_now_vs_g19}
\end{table*}

\begin{sidewaystable*}
    \centering
    \begin{tabular}{|c|c|c|c|c|c|c|c|c|c|c|c|c|}
        \hline
        Species & $\Delta$CS & $\Delta$Peri1 & $\Delta$Ap1 & $\Delta$Peri2 & $\Delta$Ap2 & $\Delta$Peri3 & $\Delta$Ap3 & $\Delta$Peri4 & $\Delta$Ap4 & $\Delta$Peri5 & $\Delta$Ap5 & Observational \\ \hline
	\ce{CO}	&	4.1(-2)	 & 	1.1(-1)	 &	1.0(-1)	 &	1.1(-1)	 &	1.1(-1)	 &	1.1(-1)	 &	1.1(-1)	 &	1.2(-1)	 &	1.2(-1)	 &	1.2(-1)	 &	1.3(-1)	&	1.2-2.3(-1)	\\	\hline
	\ce{CO2}	&	5.0(-2)	 & 	3.3(-1)	 &	3.3(-1)	 &	3.3(-1)	 &	3.3(-1)	 &	3.2(-1)	 &	3.2(-1)	 &	3.1(-1)	 &	3.1(-1)	 &	3.0(-1)	 &	3.0(-1)	&	6(-2)	\\	\hline
	\ce{CH4}	&	1.5(-1)	 & 	1.1(-2)	 &	1.1(-2)	 &	1.1(-2)	 &	1.1(-2)	 &	1.1(-2)	 &	1.1(-2)	 &	1.1(-2)	 &	1.1(-2)	 &	1.1(-2)	 &	1.1(-2)	&	1.5(-2)	\\	\hline
	\ce{H2CO}	&	4.1(-3)	 & 	1.1(-2)	 &	1.1(-2)	 &	1.1(-2)	 &	1.1(-2)	 &	1.1(-2)	 &	1.1(-2)	 &	1.2(-2)	 &	1.2(-2)	 &	1.2(-2)	 &	1.2(-2)	&	1.1(-2)	\\	\hline
	\ce{CH3OH}	&	7.3(-2)	 & 	4.1(-2)	 &	4.1(-2)	 &	4.1(-2)	 &	4.1(-2)	 &	4.1(-2)	 &	4.1(-2)	 &	4.1(-2)	 &	4.1(-2)	 &	4.2(-2)	 &	4.1(-2)	&	2.4(-2)	\\	\hline
	\ce{NH3}	&	7.0(-3)	 & 	1.1(-2)	 &	1.1(-2)	 &	1.1(-2)	 &	1.1(-2)	 &	1.1(-2)	 &	1.1(-2)	 &	1.1(-2)	 &	1.1(-2)	 &	1.1(-2)	 &	1.1(-2)	&	7(-3)	\\	\hline
	\ce{C2H2}	&	8.1(-13)	 & 	2.2(-8)	 &	2.4(-8)	 &	2.3(-8)	 &	2.2(-8)	 &	2.1(-8)	 &	1.9(-8)	 &	1.8(-8)	 &	1.6(-8)	 &	1.5(-8)	 &	1.4(-8)	&	1-3(-3)	\\	\hline
	\ce{C2H6}	&	2.5(-3)	 & 	5.5(-6)	 &	5.9(-6)	 &	5.8(-6)	 &	5.5(-6)	 &	5.1(-6)	 &	4.8(-6)	 &	4.4(-6)	 &	4.0(-6)	 &	3.7(-6)	 &	3.4(-6)	&	6(-3)	\\	\hline
	\ce{HCOOH}	&	2.9(-6)	 & 	1.5(-4)	 &	1.5(-4)	 &	1.5(-4)	 &	1.5(-4)	 &	1.6(-4)	 &	1.6(-4)	 &	1.6(-4)	 &	1.6(-4)	 &	1.6(-4)	 &	1.6(-4)	&	9(-4)	\\	\hline
	\ce{CH3CHO}	&	4.8(-5)	 & 	1.9(-5)	 &	2.0(-5)	 &	1.9(-5)	 &	1.9(-5)	 &	1.8(-5)	 &	1.7(-5)	 &	1.7(-5)	 &	1.6(-5)	 &	1.5(-5)	 &	1.5(-5)	&	2(-4)	\\	\hline
	\ce{CH3OCHO}	&	2.4(-4)	 & 	3.0(-4)	 &	3.0(-4)	 &	3.0(-4)	 &	2.9(-4)	 &	2.8(-4)	 &	2.7(-4)	 &	2.6(-4)	 &	2.5(-4)	 &	2.4(-4)	 &	2.3(-4)	&	8(-4)	\\	\hline
	\ce{CH3OCH3}	&	4.9(-3)	 & 	2.1(-4)	 &	2.2(-4)	 &	2.2(-4)	 &	2.1(-4)	 &	2.0(-4)	 &	1.9(-4)	 &	1.8(-4)	 &	1.7(-4)	 &	1.6(-4)	 &	1.5(-4)	&	$<$5(-3)	\\	\hline
	\ce{(CH2OH)2}	&	4.1(-8)	 & 	2.6(-4)	 &	2.5(-4)	 &	2.6(-4)	 &	2.5(-4)	 &	2.6(-4)	 &	2.5(-4)	 &	2.6(-4)	 &	2.6(-4)	 &	2.7(-4)	 &	2.6(-4)	&	2.5(-3)	\\	\hline
	\ce{HCN}	&	9.0(-6)	 & 	1.7(-6)	 &	1.8(-6)	 &	1.7(-6)	 &	1.7(-6)	 &	1.6(-6)	 &	1.5(-6)	 &	1.5(-6)	 &	1.4(-6)	 &	1.3(-6)	 &	1.3(-6)	&	2.5(-3)	\\	\hline
	\ce{HNC}	&	1.4(-10)	 & 	7.4(-9)	 &	7.2(-9)	 &	7.3(-9)	 &	7.5(-9)	 &	7.8(-9)	 &	8.1(-9)	 &	8.4(-9)	 &	8.7(-9)	 &	8.9(-9)	 &	9.2(-9)	&	4(-4)	\\	\hline
	\ce{HNCO}	&	8.0(-9)	 & 	2.2(-6)	 &	2.1(-6)	 &	2.1(-6)	 &	2.2(-6)	 &	2.3(-6)	 &	2.3(-6)	 &	2.4(-6)	 &	2.5(-6)	 &	2.6(-6)	 &	2.6(-6)	&	1(-3)	\\	\hline
	\ce{CH3CN}	&	1.1(-9)	 & 	4.1(-14)	 &	5.3(-14)	 &	4.5(-14)	 &	4.3(-14)	 &	4.1(-14)	 &	3.2(-14)	 &	3.3(-14)	 &	2.1(-14)	 &	2.6(-14)	 &	2.1(-14)	&	2(-4)	\\	\hline
	\ce{HC3N}	&	1.5(-20)	 & 	2.5(-18)	 &	2.4(-18)	 &	2.4(-18)	 &	2.3(-18)	 &	5.7(-18)	 &	2.1(-18)	 &	4.9(-18)	 &	1.9(-18)	 &	4.5(-18)	 &	1.7(-18)	&	2(-4)	\\	\hline
	\ce{NH2CHO}	&	1.6(-10)	 & 	5.1(-7)	 &	4.4(-7)	 &	4.5(-7)	 &	4.5(-7)	 &	4.7(-7)	 &	4.7(-7)	 &	5.0(-7)	 &	4.9(-7)	 &	5.3(-7)	 &	5.2(-7)	&	1.5(-4)	\\	\hline

    \end{tabular}
    \caption{The amount of material lost to the gas phase, with respect to the amount of water lost, for selected chemical species at various times/intervals in the models. Values are indicated in the form $A(B) = A \times 10^B$. `$\Delta$CS' corresponds to material the lost during the cold storage phase (4.5~Gyr); `$\Delta$Peri' corresponds to the loss from the previous aphelion to the current perihelion, and `$\Delta$Ap' the loss from the previous perihelion to the current aphelion, with the orbit number also indicated.}
    \label{hb_delta_gas}
\end{sidewaystable*}

\section{Conclusions}
\label{conclusions}

The model presented herein uses the first solid-phase chemical kinetics model adapted for cometary ices and adds in the capability of modeling the active phase as well as the earlier period of ``cold storage'' in the outer solar system. Previously, the model was limited to a constant set of physical conditions but is now capable of modeling changes in temperature due to changing radial distance from the sun as well as ice depth. The model also includes a careful treatment for the effects of back-diffusion on H and H$_2$ mobility within the ice, as well as incorporating a selection of new nondiffusive processes based on recent interstellar-ice treatments. With these new adaptations, the model results may be more directly matched with observed gas-phase abundances from active-phase comets.

The application of the model firstly to 10~K cold storage conditions, representative of the Oort Cloud, produces a fairly rapid ($\sim$1000~yr) and rich chemistry in the upper $\sim$1~micron of the ice--dust mixture, caused primarily by interstellar-UV processing, as found by G19. In the deeper ice layers, to depths on the order of 10~m, Galactic cosmic-ray processing inducing a similar, but much slower chemistry that allows complex organic molecules (COMs) to build up in modest abundances. During the 4.5~Gyr of interstellar UV and GCR processing, around 19~cm of ice material is lost to space, leading to the concentration of the dust component down to a similar depth. Although the chemically rich upper micron of material is gradually lost over and over again, the chemistry is fast enough to maintain a fairly steady profile of molecular abundances over time.

The active-phase models, tailored for comet Hale--Bopp, consider five identical active-phase orbits immediately after exiting the cold storage phase. The orbital parameters for each orbit are based on the most recent observed orbit for this comet. Each solar approach phase leads to the loss of approximately 19~cm of ice from the surface, purely through sublimation over the whole comet surface (not localized outbursts). The modeled solid-phase chemistry is found to stabilize after two full orbits, such that the chemical composition as a function of depth is very similar when comparing the same point in each orbit. After this second orbit, chemical changes are much more gradual with time and trends are much more consistent.

Thus, the first solar approach is the most chemically unique, as it corresponds to the period when the outer layers of material, processed by UV during cold storage, is lost to space. Deeper material, processed by cosmic rays over billions of years, then becomes the de facto surface, which is itself then processed by UV in the upper micron of material. Further orbits release more surface material to reveal yet deeper ices. Within the five orbits modeled here, a substantial amount of material remains from the period of GCR processing. Although the chemical profiles remain qualitatively similar between orbits 2--5, dust concentration continues to greater depths each time.

As the chemistry has stabilized by this point and changes over time appear to be minimal, we can take the profile at any given point in the orbit and easily extrapolate to a larger number of orbits than the five performed with this model. Subsequent orbits should maintain comparable profiles at slightly reduced fractional abundances of non-pristine ice.

Compared to observations of comet Hale--Bopp, our fractional abundances of ``pristine'' ice (i.e. CO, \ce{CO2}, \ce{CH4}, \ce{NH3}, \ce{H2CO} and \ce{CH3OH}) as a fraction of water all match within an order of magnitude. However, all non-pristine ice species (i.e. product species) underestimate observations by at least one order of magnitude. Most remain within four orders of magnitude, but certain cases can underestimate observations by as much as fifteen orders of magnitude. In particular, the amino acid glycine is still produced in low abundance, below many other COMs. As a result, the current model and conditions still appear insufficient to replicate observed complexity within cometary ices. 

It therefore seems essential to include an initially richer ice composition, to account for inherited primordial material. This supports work by \citet{Drozdovskaya2019}, who found similar results for comet 67P in which the composition had similarities with IRAS 16293 abundances for COMs and other molecules. More specific comet simulations based on 67P conditions, including the effects of differing initial ice compositions, would be necessary to confirm the significance of these findings to the present modeling treatment. We anticipate such a study in a follow-up paper. In such a context, not only the formation of COMs and other species but the destruction and/or survival of the same, through cold storage and ultimately solar approach, must also be considered.

\section*{Acknowledgement}
This work was funded by the NASA Emerging Worlds Program, grant no. NNX17AE23G. DAC thanks the NASA FINESST program for a graduate fellowship. We thank the IAU Minor Planet Center for providing the orbital data necessary to run our models. We also thank our fellow research group member K. Stelmach for critical readings of the manuscript. We also thank our referees for their helpful comments in improving this paper.

\appendix

\section{Kinetic Monte Carlo simulations of reaction rates in bulk-ice layers}\label{appendixA}

Kinetic Monte Carlo simulations of diffusive reactions were carried out to determine the degree of back-diffusion affecting bulk-ice reactions of H or H$_2$. The model is broadly similar to that described in Section~\ref{MCsims}, except that when diffusers meet (i.e.~hop into the same site) they are assumed to react, ending that particular simulation and triggering the next to begin. These simulations may be considered to be specifically representative of the H + H $\rightarrow$ H$_2$ reaction system.

The simulations must continue until a reaction occurs, even if a diffuser has attempted to cross an upper or lower boundary of the ice layer. Thus, when a diffuser does cross a boundary, an action must be taken to conserve the total number of diffusers in the ice. Two separate sets of simulations were run to take account of these cases in different ways: (i) if a boundary is crossed, then that hop is ignored, the particle is replaced in its previous position, and it is allowed to move again in a different, random direction; (ii) the ice is treated as fully periodic, so that the crossing of a boundary places the diffuser into the other end of the ice layer, which could lead directly to a reaction. Case (i) may be considered to be closer to the interstellar dust-grain situation, in which the lower bound of the ice is blocked by the grain itself. Case (ii) is closer to the picture presented in the comet model, in which a diffuser may leave the ice layer in question (while preserving the total number of diffusers), while diffusers might also enter from another adjacent ice layer.

\begin{figure}
    \centering
    \includegraphics[width=0.45\textwidth]{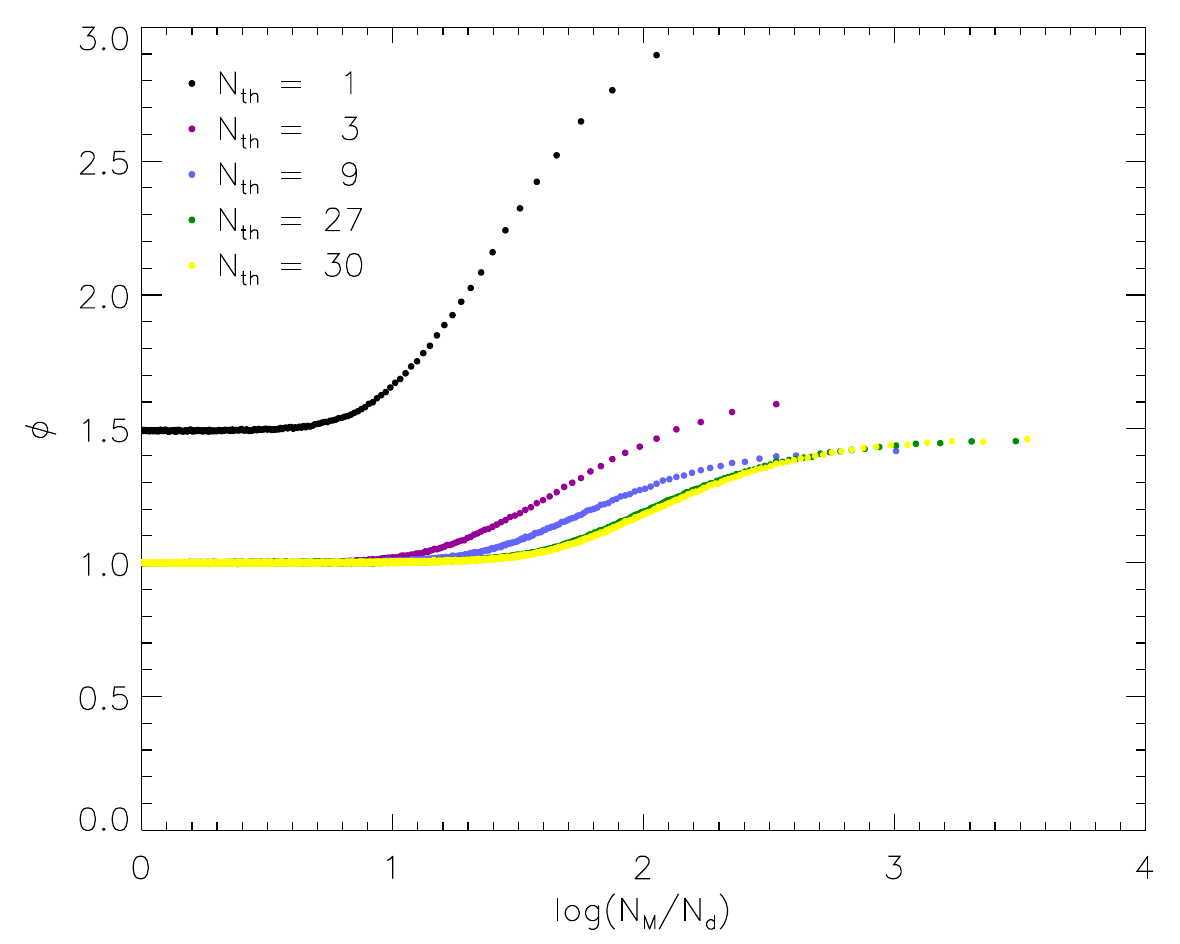}
    \caption{Back-diffusion factor, $\phi$, affecting the rates of diffusive reactions in bulk-ice layers of various thicknesses, $N_{\mathrm{th}}$, based on kinetic Carlo Models with full periodic boundary conditions. Results are shown as a function of the degree of occupation of sites ($N_{\mathrm{M}}/N_{\mathrm{d}}$). Details of the models are described here and in Section~\ref{MCsims}.}
    \label{appendix1}
\end{figure}

Fig.~\ref{appendix1} shows the reactive back-diffusion factor, $\phi$, as determined for case (ii), as a function of the inverse of the degree of occupation of sites ($N_{\mathrm{M}}/N_{\mathrm{d}}$), following the same method of representation as used by \cite{Willis2017}. Occupation is highest toward the left, while the rightmost point for each color indicates the case where only two diffusers/reactants are present in that ice layer. Different colors show the results for different ice-layer thicknesses, with values chosen to coincide with the layer thicknesses actually used in the comet model (except for $N_{\mathrm{th}}=30$, which is simply the thickest ice tested). For all thicknesses other than $N_{\mathrm{th}}=1$ and $3$, the back-diffusion factor ranges between 1 and $\sim$1.5, with the $N_{\mathrm{th}}=3$ models also coming very close to this behavior. The true exception is $N_{\mathrm{th}}=1$, which corresponds to a single layer in which any upward or downward diffusion is technically counted as a hop, while not leading to an actual change of position. This has the effect of artificially inflating the number of hops by a factor of 1.5 ($=6/4$) for all $N_{\mathrm{M}}/N_{\mathrm{d}}$ values.

\begin{figure}
    \centering
    \includegraphics[width=0.45\textwidth]{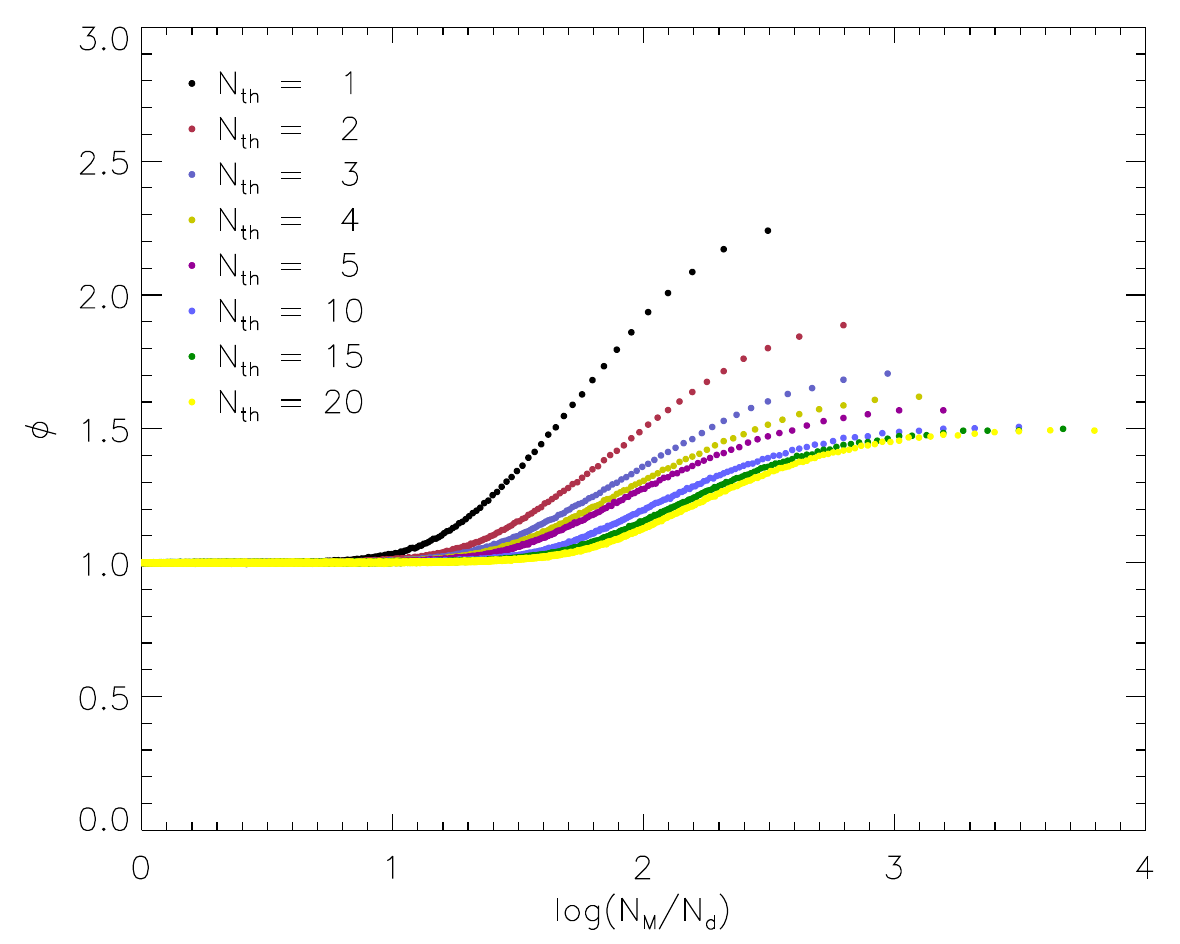}
    \caption{As Fig.~\ref{appendix1}, but using a model in which boundary-crossing is not allowed and is not counted in the total number of hops.}
    \label{appendix2}
\end{figure}

Fig.~\ref{appendix2} shows the results for case (i), in which no diffusion across the boundaries is allowed, using a slightly wider ice ($N_{\mathrm{lat}}=25$) and a slightly lower maximum thickness ($N_{\mathrm{th}}=20$). For $N_{\mathrm{th}}=1$ the anomalous factor of 1.5 is removed, while the greatest back-diffusion factor -- corresponding, as usual, to the presence of two diffusers/reactants -- is still fairly small at around 2.25. The particular case of two reactants on a surface was studied in more detail by \cite{Willis2017}, and they derived an expression for $\phi$ that was dependent on the number of sites, $N_{\mathrm{M}}$ (see their Table 2). Their expression for a flat surface morphology returns a value $\phi = 2.23$ for $N_{\mathrm{lat}}=25$, matching the data shown in Fig.~\ref{appendix2}. The same expression (multiplied by a factor 1.5) may also be used to reproduce the most extreme $\phi$-value for $N_{\mathrm{th}}=1$ shown in Fig.~\ref{appendix1}.

The same expression may be used to extrapolate the most extreme result of the simulations to a surface of a size appropriate to a comet, as indeed is done in the comet model to determine the back-diffusion factor for the surface reaction rates. Adopting a (square) comet surface of 1~km $\times$ 1~km, a total of $\phi \simeq 18$ would be expected. While this is not insignificant, it represents a value even higher than would be achieved in the first bulk-ice layer ($N_{\mathrm{th}}=3$), as judged by Fig.~\ref{appendix1}, while the $N_{\mathrm{th}}=9$ layer beneath it would be expected to converge to values in the range 1 -- 1.5 for all $N_{\mathrm{M}}/N_{\mathrm{d}}$ values. These $\phi$-values are all minuscule compared with the back-diffusion factor for transfer between layers (instead of reaction).

Thus, while the Monte Carlo simulations cannot be used to calculate directly the degree of three-dimensional back-diffusion affecting diffusive reaction rates in bulk-ice layers of sizes appropriate to comets (i.e.~$N_{\mathrm{lat}}$ >> 25), the limits placed on the most extreme cases indicate that the effect is likely small enough to ignore. Judging by the results for $N_{\mathrm{th}}>1$ in Fig.~\ref{appendix2}, the effect of back-diffusion on bulk-ice reaction rates in the {\it interstellar-grain} case is also more or less negligible, and will always be substantially less than the factor of $\sim$5 found for surface reactions ($N_{\mathrm{th}}=1$), both here and in \cite{Willis2017}.

\printcredits

\bibliographystyle{cas-model2-names}

\bibliography{cas-refs}

\end{document}